# Realizing Linear Synaptic Plasticity in Electric Double Layer-Gated Transistors for Improved Predictive Accuracy and Efficiency in Neuromorphic Computing

*Nithil Harris Manimaran[1], Cori Sutton[2], Jake Streamer[3], Cory Merkel[4], and Ke Xu\*[1,2,5]*

[1]Microsystems Engineering,

[2]School of Physics and Astronomy,

[3]Multidisciplinary Study,

[4]Department of Electrical and Computer Engineering,

[5]School of Chemistry and Material Science,

Rochester Institute of Technology, Rochester, NY, 14623, USA

Correspondence and requests for materials should be addressed to K.X. (email:ke.xu@rit.edu).

**ABSTRACT**

Neuromorphic computing offers a low-power, parallel alternative to traditional von Neumann architectures by addressing the sequential data processing bottlenecks. Electric double layer-gated transistors (EDLTs) resemble biological synapses with their ionic response and offer low power operations, making them suitable for neuromorphic applications. A critical consideration for artificial neural networks (ANNs) is achieving linear and symmetric plasticity (i.e., weight updates) during training, as this directly affects accuracy and efficiency. This study uses finite element modeling to explore EDLTs as artificial synapses in ANNs and investigates the underlying mechanisms behind the nonlinear weight updates observed experimentally in previous studies. By solving modified Poisson-Nernst-Planck (mPNP) equations, we examined ion dynamics within an EDL capacitor and their effects on plasticity, revealing that the rates of EDL formation and dissipation are concentration-dependent. Fixed-magnitude pulse inputs result in decreased formation and increased dissipation rates, leading to nonlinear weight updates. For a pulse magnitude of 1 V, both 1 ms 500 Hz and 5 ms 100 Hz pulse inputs saturated at less than half of the steady state EDL concentration, limiting the number of accessible states and operating range of devices. To address this, we developed a predictive linear ionic weight update solver (LIWUS) in Python to predict voltage pulse inputs that achieve linear plasticity. We then evaluated an ANN with linear and nonlinear weight updates on the MNIST classification task. The ANN with LIWUS-provided linear weight updates required 19% fewer (i.e. 5) epochs in both training and validation than the network with nonlinear weight updates to reach optimal performance. It achieved a 97.6% recognition accuracy, 1.5 – 4.2% higher than with nonlinear updates, and a low standard deviation of 0.02%. The network model is amenable to future spiking neural network applications, and the performance improvements with linear weight update is expected to increase for complex networks with multiple hidden layers.

**INTRODUCTION**

Most present-day computing hardware is based on the von Neumann architecture where both data and instructions are stored in the memory and are fetched and executed by the CPU, one at a time and in order. This sequential processing leads to the so-called von Neumann bottleneck where computations are limited due to the shuttling of information back and forth between the memory and CPU.[1] Domain specific architectures have emerged over the past few decades to achieve better performance-to-cost ratios.[2-4] However, there is a necessity for a low-power, highly-parallel architecture to overcome this bottleneck, and neuromorphic computing aims to address this challenge.[5,6] Artificial neural networks (ANNs) are brain-inspired parallel systems that require less power than traditional von Neumann computing systems.[7-10] The computation and memory units reside together, improving device speed. For example, SpiNNaker, a supercomputer developed by the Advanced Processor Technologies (APT) Research Group at the University of Manchester has 18 cores per die, with each core emulating ~1000 neurons.[11] Here, the processing layers, memory layers, and routers (communication layers) are all integrated. Systems such as SpiNNaker fall under a specialized subset of ANNs called spiking neural networks (SNNs). The layers or components of each layer are event-driven like as observed in biological systems[12,13], meaning they participate in computation (and consume power) only when triggered and reacting to the previous layer, making the system suitable for low power applications. Concurrently, by mimicking the brain's parallelism and spatiotemporal processing, neuromorphic computing systems can dynamically respond in real time. In applications such as autonomous vehicles and robots, this capability enables instantaneous feedback, enhancing pattern recognition, navigation, decision-making, and ensures low latency for responses that are crucial for performance and safety.[14,15]

Neural networks seek to emulate the brain's functionality and the transmission of information between neurons. Synapses are the junctions between neurons, with each neuron having up to thousands of synapses facilitating communication with other neurons.[16] A synapse is illustrated in Fig. 1 (a), where information is transmitted from the presynaptic neuron to the postsynaptic neuron, in the form of neurotransmitters such glutamate or $\gamma$-aminobutyric acid (GABA). Neurotransmitters diffuse across the synaptic cleft (20 – 50 nm wide) and bind with ligand-gated ion channels, which allow ions such as sodium or potassium to move into or out of the postsynaptic neuron, modifying the potential of the cell membrane; if membrane potential increases and surpasses the neuron's threshold, then the neuron will produce an action potential.[17] Example action potentials are depicted in Fig. 1 (b). As positively charged sodium ions start to accumulate inside of the cell in response to a presynaptic stimulus, the membrane potential slowly increases from the resting potential of -70 mV to the threshold of -55 mV. As the threshold is reached, a positive feedback loop is created through voltage-gated ion channels, causing a flood of positive ions to move into the cell, creating a +40 mV voltage spike. Then, the potential sharply decreases back to the resting potential due to a combination of sodium ion influx ceasing and delayed positively charged potassium efflux.[18] Fig. 1 (b) also depicts a case where the threshold potential is not reached, resulting in the neuron not firing. Note that each spike is voltage or event driven and lasts 1 – 5 ms, depending on the cell type and size.[18,19] When a series of stimuli occur or the neuron receives a train of action potentials, subsequent spikes do not allow the potential to relax immediately back to the resting potential and the overall potential increases (Fig. 1 (c)). This form of temporal memory enables efficient spatiotemporal encoding of information in the brain. Moreover, the strength of a neuron's response to a presynaptic spike can change over time. This is known as *synaptic plasticity* and is a major component of biological learning; a positive change (potentiation) strengthens synapses while a negative change (depression) weakens them. One of the drivers for these synaptic changes is the relative timing of pre- and postsynaptic spikes (spike-timing-dependent plasticity, or STDP), which allows the brain to efficiently learn causal relationships.[20,21]

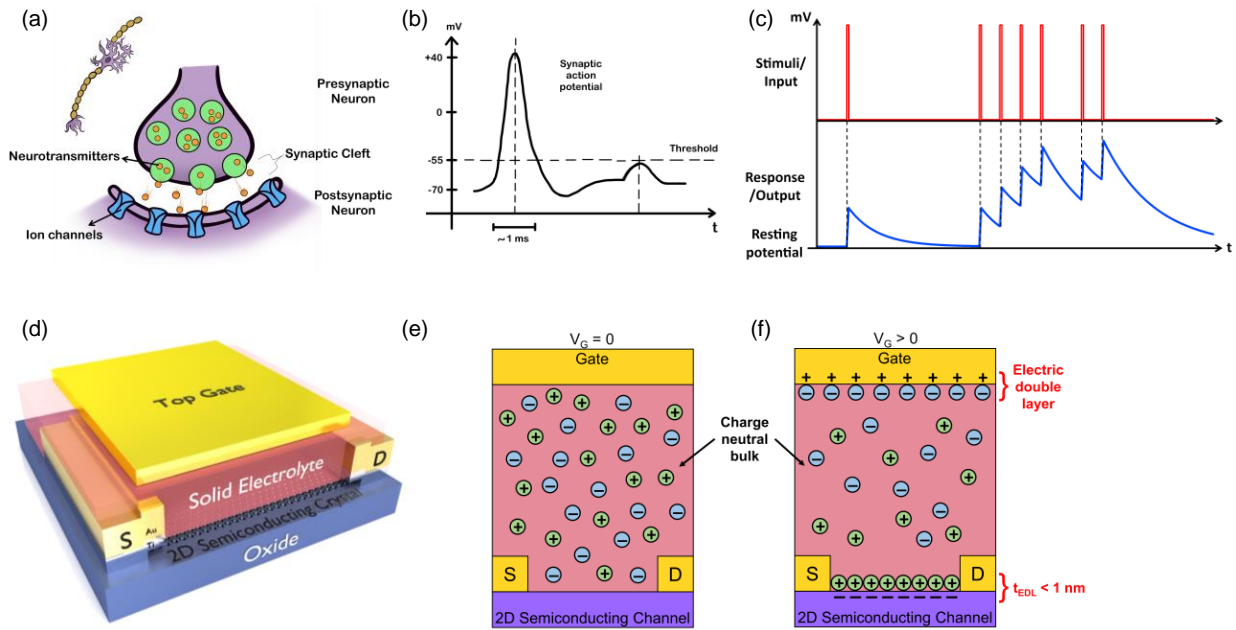

**Figure 1**: Illustrations of (a) a synapse, (b) example action potentials, and (c) synaptic plasticity. (d) Schematic of a top gated EDLT and corresponding ion distributions under (e) no gate bias and (f) a positive gate bias.

A diverse array of materials and device architectures has been proposed to physically emulate synaptic functions. This includes two-terminal devices such as resistive random access memory (ReRAM)[22–24] and phase change memory (PCM)[25], and three-terminal devices like ferroelectric FETs (FeFETs)[26] and electric double layer transistors (EDLTs).[27–29] The inclusion of an additional gate terminal in three-terminal devices facilitates more precise control over resistance and permits simultaneous information reception and readout. Among three-terminal devices, the EDLT has attracted considerable interest for its resemblance to biological synapses and its capability to facilitate STDP and complex synaptic behaviors through temporal and spatial integration.[27,30–36] A schematic of EDLT is shown in Fig. 1 (d). The top gate functions as the presynaptic neuron with the gate voltage ($V_G$) acting as input stimuli, the electrolyte would take the role of neurotransmitters and the postsynaptic current would be the source-drain current ($I_D$). Under an applied gate voltage (input stimuli), mobile ions in the electrolyte respond to the electric field, forming electric double layers at the gate-electrolyte and electrolyte-channel interfaces, thereby altering the channel resistance (Fig. 1 (e) & (f)). Irrespective of the thickness of electrolyte used, the thickness of the EDL is always ~1 nm, resulting in capacitance densities up to 10 µF/cm$^2$ and charge carrier concentrations in the order of 10$^{14}$ cm$^{-2}$,[37–40] which are suitable for low power and neuromorphic applications. Furthermore, the ionic (and electrical) response of an EDLT resembles that of a biological synapse, showing potential in developing synaptic devices.[35,36,41,42] A recent study by Min and Cho demonstrated both short and long-term potentiation in a chitosan electrolyte – Ta$_2$O$_5$ hybrid EDLT, and also reported CMOS compatibility, on/off ratio in the order of 10$^7$, and a capacitance density of ~0.2 µF/cm$^2$ at 100 Hz.[43]

A key aspect of designing ANNs is ensuring a linear and symmetric weight update, which decreases noise, enhances prediction power and accuracy, and reduces the training and computational demands of the network.[10,44–46] The different conductance states that could be obtained with the EDLTs are analogous to biological synaptic weights. Neural networks based on EDLTs can be trained more efficiently if the conductance states increase (potentiate) or decrease (depress) linearly in response to input pulses, and if there is symmetry between potentiation and depression (Fig. 2 (a)). Although EDLTs emulate biological synapses and operate at low voltages, their synaptic weight updates are inherently nonlinear (Fig. 2 (a)), as shown in other studies.[29,35,36,47–51] Therefore, it is crucial to: (1) understand the fundamental mechanisms behind this nonlinear behavior, and (2) discover potential strategies to achieve linear plasticity, which will enhance the performance of neural networks.

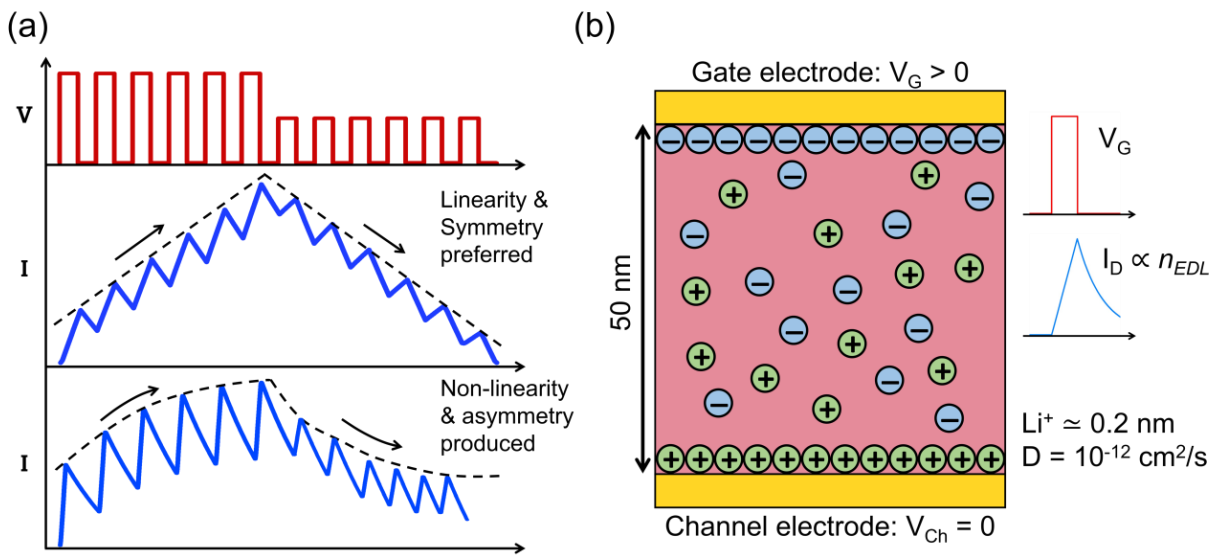

**Figure 2**: (a) An example illustration depicting a linear weight update preferred by ANNs vs a nonlinear weight update obtained in EDLTs, both in response to linear input pulses. (b) Schematic of the EDLT's capacitive region modeled in COMSOL as a 50 nm x 50 nm parallel plate capacitor, under a positive gate bias.

In this study, we investigate the EDL ion dynamics through finite element modeling via COMSOL Multiphysics and develop a predictive model in Python to simulate necessary pulse inputs for a linear weight update. Both linear and nonlinear weight updates from COMSOL are applied to a custom multi-layer perceptron (MLP) ANN to evaluate their performance in supervised learning and image classification. In COMSOL, EDL ion dynamics was modeled by solving the modified Poisson-Nernst-Planck (mPNP) equations. 1 V pulse trains of varying frequencies show the EDL ion density ($n_{EDL}$) saturating within a few pulses (e.g., 5 ms pulses at 100 Hz saturated after 7 pulses, with $n_{EDL}$ saturating at 47% of the magnitude observed at steady state). We found that both the formation and dissipation rates of EDL depend on the instantaneous EDL concentration: as the concentration increases, the formation rate decreases while the dissipation rate increases, leading to saturation and nonlinear weight update. To achieve linear plasticity, a predictive model was developed in Python, trained on steady state EDL data, to predict the necessary pulse inputs for a linear weight update. Using both linear and nonlinear weight updates simulated in COMSOL, an ANN was designed using TensorFlow for image classification and demonstrated a 1.5 – 4.2% increase in recognition accuracy with linear updates, with further improvements expected for more complex networks with additional layers and nodes.

**RESULTS AND DISCUSSION**

We first studied the relationship between frequency and EDL concentration using the parallel plate capacitor geometry shown in Fig. 2 (b). In previous studies, we have demonstrated that the modelled EDL ion densities in this geometry accurately track the voltage dependence of EDLT charge carrier densities measured experimentally by Hall effect, with the model predicting about 4 to 10 times higher densities than experiment.[38,52] 1 V pulses at two different pulse widths ($w$, 1 ms and 5 ms) were applied. The 5 ms input pulses with frequencies ranging from 100 Hz to 10 Hz are shown in Fig. 3 (a) and their corresponding responses in Fig. 3 (b), while Fig. 3 (c) shows the responses to 1 ms pulses with frequencies from 500 Hz to 10 Hz. As a reference, a 500 ms step voltage of 1 V was also simulated to track the steady state EDL concentration at 1V. Based on our previous experimental studies, this timescale is sufficient to allow EDL to fully saturate at this applied voltage and reach steady state.[42] The EDL ion concentration saturated at a value of $1.91 \times 10^{14}$ ions/cm$^2$ ($n_{EDL}^{SS}$), representing the maximum possible ion density at an input of 1 V. Upon voltage removal, EDL concentration gradually decreased, reaching the initial resting state EDL concentration in approximately 100 ms.

The 5 ms pulses are discussed first. As subsequent voltages were applied before the ions could revert to their initial state, an increase in EDL concentration was observed after each pulse, demonstrating synaptic facilitation. However, similar to what was observed in previous experimental studies, the increase in EDL concentration was not linear in response to fixed input signals,[29,35,36,50] and the EDL concentration saturated after just a few pulses, as shown in Fig. 3 (b). A similar trend depicting depression is shown in Fig. S1. This saturation concentration was significantly lower than the steady state EDL concentration, with 100 Hz saturating at $9 \times 10^{13}$ ions/cm$^2$, over 2 times less than the steady state EDL concentration of $1.91 \times 10^{14}$ ions/cm$^2$. No consecutive EDL accumulation was observed with the 10 Hz pulses because the interval between pulses is longer than ions' dissipation time, and the ions reverted to their resting state before the subsequent pulse was applied.

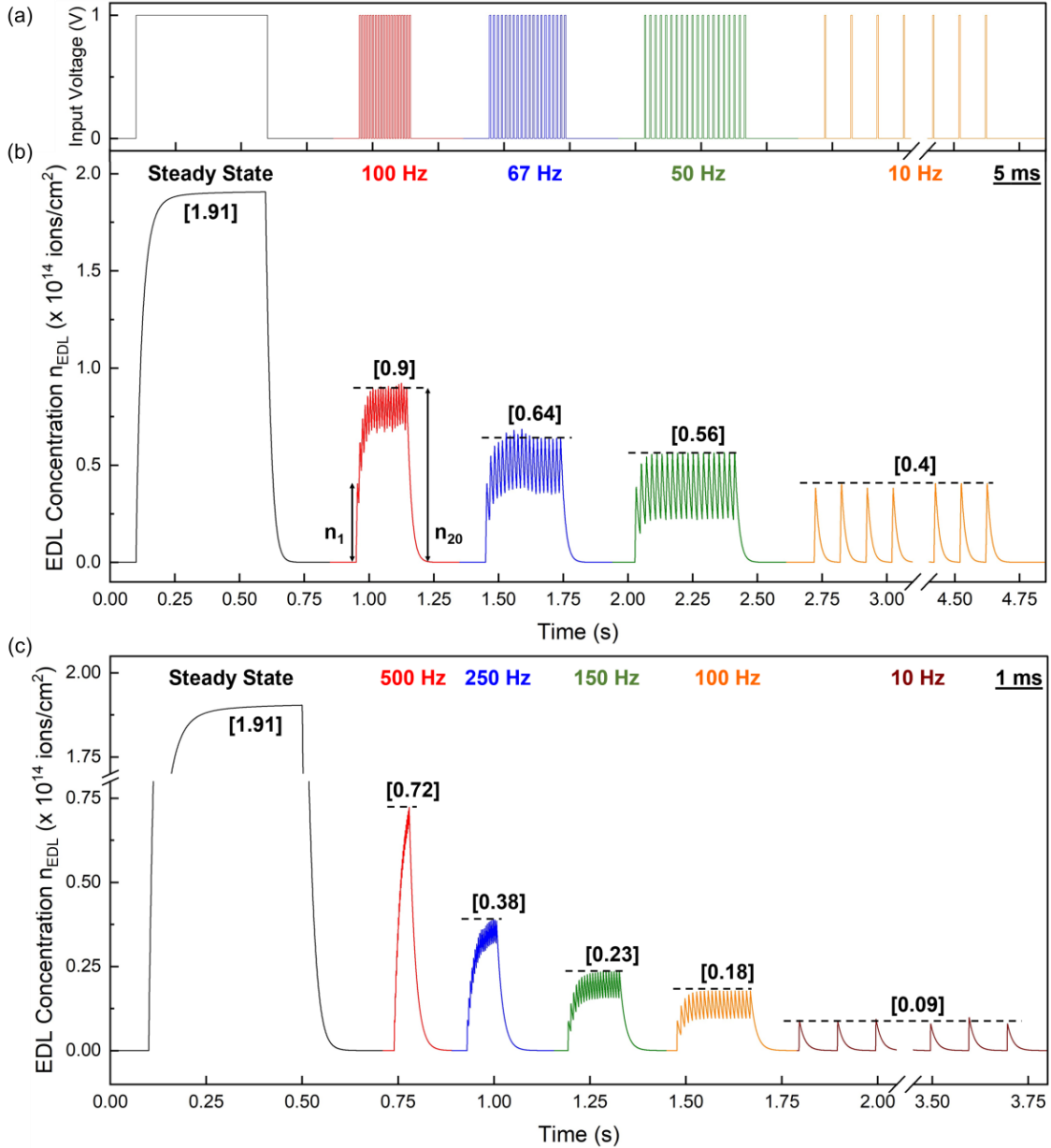

**Figure 3:** (a) 1 V input voltages to the COMSOL model; a step voltage to simulate steady state followed by 5 ms pulses of 100 Hz, 67 Hz, 50 Hz, & 10 Hz, and (b) their corresponding output EDL concentrations. (c) EDL concentrations in response to 1 ms pulses of frequencies 500 Hz – 10 Hz.

As discussed in the introduction, a linear synaptic weight update, which corresponds to a linear increase in EDL concentration in our modeling, is beneficial for increasing training accuracies and reducing noises in backpropagating neural networks.[10,44–46] Importantly, each weight update must also lead to distinguishable states. As the frequency increases from 10 Hz to 100 Hz, the number of distinguishable states increases from 1 to 7. Furthermore, the difference in magnitudes between the first and last distinguishable states ($\Delta n$) also increased with frequency. For example, $\Delta n = 2.34 \times 10^{13}$ ions/cm$^2$ for 67 Hz and $4.94 \times 10^{13}$ ions/cm$^2$ for 100 Hz. Since the modelled EDL ion densities have proven to correlate with the experimentally measured charge carrier densities in EDLTs,[38,52] the simulated $\Delta n$ directly relates to EDLTs' on/off ratio and the range over which the synaptic weight changes, which has significant impact on the neural network's performance, training time and power consumption. A larger range results in a better distinction of on/off states by the neural network, reduces test-to-test variability, and reduces noise propagation which could cause unwanted power consumption.[53,54]

Although the saturation EDL concentration increases with frequency, at 100 Hz it still saturates at approximately half the value of the steady state. Similarly, even though the number of distinct states increases with frequency, 7 distinct linear states (at 100 Hz) is insufficient for practical applications. Note that 5 ms pulses were simulated up to 100 Hz to allow pulse period $T \geq 2w$. By decreasing the pulse width to 1 ms, while maintaining this condition, frequencies up to 500 Hz could be simulated (Fig. 3 (c)). With $w = 1$ ms, the number of distinct states increased to 20 for 500 Hz but decreased to 5 for 100 Hz. Moreover, $\Delta n$ decreased to $0.86 \times 10^{13}$ ions/cm$^2$ for 100 Hz (~6 times smaller than the 5 ms 100 Hz) and was $6.32 \times 10^{13}$ ions/cm$^2$ for 500 Hz. These results are summarized in Table 1, and the 100 Hz and 500 Hz results are highlighted. Although the number of states increased for the 500 Hz pulse train, the EDL density at saturation was smaller compared to the 5 ms responses, implying a smaller sheet carrier density (or drain current) in EDLTs.[38] The effect of frequency on the saturation EDL concentration for different pulse widths between 1 ms and 5 ms are highlighted in Fig. S2. To sum up, increasing the frequency of pulses while keeping the pulse width constant shows an increase in number of the distinct states and the saturation EDL concentration attained by the pulses ($n_{EDL}^P$). On the other hand, increasing the pulse width while the frequency is kept constant shows an increase in the number of distinct states, $n_{EDL}^P$, and $\Delta n$. However, two problems remain: 1) the saturation EDL concentration attained by the pulses ($n_{EDL}^P$) are significantly smaller than the steady state EDL concentration ($n_{EDL}^{SS}$), as summarized in Table 1; and 2) the EDL concentration increase nonlinearly, which is in agreement with experimental studies and could reduce the efficiency of neural networks.

*Table 1: The number of distinct states and usable concentration range extracted from the COMSOL simulations*

| Pulse width $w$ (ms) | Frequency (Hz) | No. of distinct states | Saturation EDL Concentration $n_{EDL}^P$ (x $10^{14}$ ions/cm$^2$) | $n_{EDL}^P/n_{EDL}^{SS}$ (%) | $\Delta n$ (x $10^{13}$ ions/cm$^2$) |
|---|---|---|---|---|---|
| 5 | 10 | 1 | 0.4 | 20.94 | 0 |
|  | 50 | 3 | 0.56 | 29.32 | 1.52 |
|  | 67 | 3 | 0.64 | 33.51 | 2.34 |
|  | 100 | 7 | 0.9 | 47.12 | 4.94 |
| 1 | 10 | 1 | 0.09 | 4.71 | 0 |
|  | 100 | 5 | 0.18 | 9.42 | 0.86 |
|  | 150 | 8 | 0.23 | 12.04 | 1.42 |
|  | 250 | 11 | 0.38 | 19.9 | 2.92 |
|  | 500 | 20 | 0.72 | 37.7 | 6.32 |

To understand the fundamental cause of the nonlinear increase (or decrease if pulsing in the depression region) of EDL ion concentration, we looked at the rate of EDL formation and dissipation. The steady state result with a step voltage of 1 V was repeated in Fig. 4 (a) to highlight the red and blue lines corresponding to EDL formation and dissipation, respectively. The corresponding rates ($dn_{EDL}/dt$ or $dc_+/dt$) of both EDL formation (red) and dissipation (blue) are shown in Fig. 4 (b). It's important to note here that the dashed lines depict the EDL rates with respect to *time*, denoted by the top x-axis; while the solid lines depict the rates with respect to *EDL concentration*, represented by the bottom x-axis. With respect to time, the rate of EDL formation increases sharply as a voltage is applied and then decays slowly, mirroring the behavior observed in the dissolution rate when the voltage is removed. This behavior is expected and explained in previous experimental and simulation studies on EDL dynamics.[34,42,55,56] Briefly, during EDL formation, as the potential drop across the EDL intensifies, the driving force for ion movement ($\Delta V$) diminishes, leading to a deceleration in the formation rate. Analogously, in the dissolution phase, the diminishing concentration gradient between the EDL and the bulk electrolyte contributes to a gradual reduction in the dissolution rate over time.

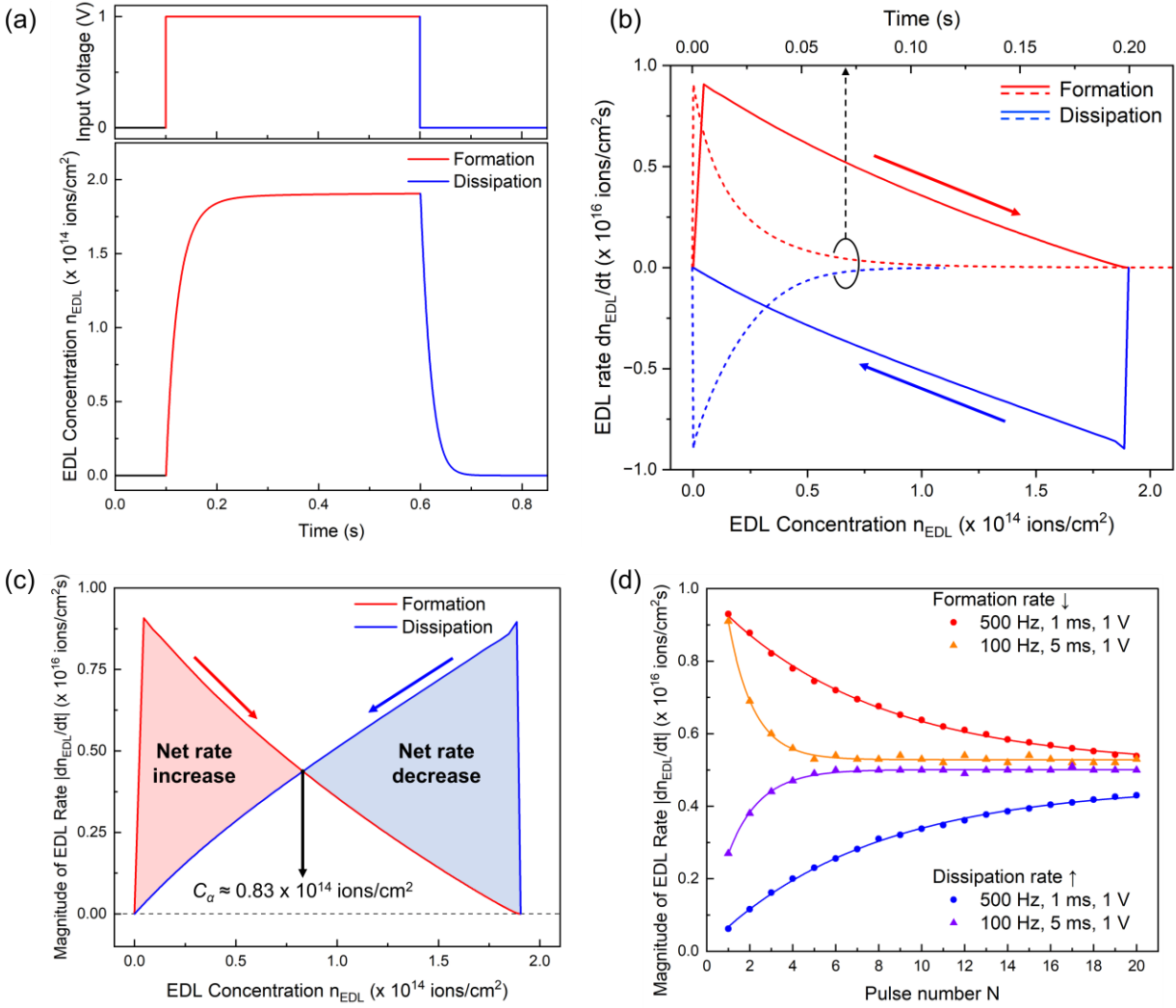

**Figure 4:** (a) A single step voltage of 1 V and its corresponding response simulated in COMSOL, (b) EDL rate (derivative of plot (a)) plotted against time (top x-axis) and instantaneous EDL concentration (bottom x-axis), (c) magnitude only of EDL formation and dissipation rates, depicting net rate of zero ($C_\alpha$), and (d) magnitude of EDL rates showing formation rate decreasing and dissipation rate increasing with pulses (1 ms at 500 Hz & 5 ms at 100 Hz), and eventually reaching an equilibrium.

The temporal dynamics of EDL rates can additionally be contextualized by considering the instantaneous EDL concentration (Fig. 4 (b) bottom x-axis), which is useful in explaining the nonlinear potentiation shown in Fig. 3. To facilitate a comparison, Fig. 4 (c) illustrates the *magnitude* of EDL formation/dissipation rate as a function of concentration. It can be observed that the rates of formation and dissipation become equal at an EDL concentration of $C_\alpha \simeq 0.83 \times 10^{14}$ ions/cm$^2$. Note that this $C_\alpha$ value would change with applied voltage, frequency, and initial bulk ion concentration (assumed to be 1 mol/L in this study). Considering a 50% duty cycle — where the duration of EDL formation equals that of its dissipation — at EDL concentrations lower than $C_\alpha$, the magnitude of formation rate is greater than that of dissipation, leading to a cumulative increase in EDL concentration after each pulse cycle. However, as the EDL concentration escalates beyond $C_\alpha$, the dissipation rate exceeds the formation rate, resulting in a net decrease in EDL concentration. Eventually, an equilibrium is attained where the EDL concentration oscillates around the $C_\alpha$ value. The observed value of $C_\alpha$ in Fig. 4 (c) is consistent with the saturation level of $0.8 \pm 0.05$ ions/cm$^2$ demonstrated in the 5 ms, 100 Hz dataset of Fig. 3 (b), which operates under a 50% duty cycle. When the duty cycle is below 50% (for instance, at frequencies lower than 100 Hz as shown in Fig. 3 (b)), the dissipation rate still follows Fig. 4 (c) but the dissipation *time* is prolonged, resulting in a lower saturation $C_\alpha$. Conversely, $C_\alpha$ is higher when duty cycle exceeds 50%. This pattern of saturation is consistent across varying pulse frequencies and widths. Fig. 4 (d) presents the rates of EDL formation and dissipation post each pulse for 500 Hz, 1 ms pulses, and for 100 Hz, 5 ms pulses. In summary, with a constant pulse magnitude, the rate of EDL formation diminishes as the EDL builds up while dissipation rate of EDL intensifies, leading to a nonlinear increase in EDL concentration until saturation is achieved.

To overcome EDL saturation and attain more linear states, it is necessary to increase either the pulse frequency, pulse width, or pulse magnitude as pulses progress. Changing the pulse frequency has been discussed above and has its limitations. Increasing the pulse width is also not ideal because it raises power consumption and limits the number of pulses (and consequently the number of states) that can be achieved before the EDL reaches full saturation (steady state). To achieve linear synaptic potentiation in EDLTs, one possible solution is to increase the pulse magnitude with each pulse. Similar approaches are used in the memory industry; for instance, incremental step pulse programming (ISPP) schemes employ this technique when testing modern memory devices. Adaptive ISPP schemes with fixed pulse widths have demonstrated improved reliability and lower power consumption.[57–59] Ferroelectric FETs used as synaptic devices have also shown improvements in accuracy with incremental pulsing schemes.[60]

However, changing the pulse magnitude to achieve a linear synaptic weight update requires calculating the magnitude of each consecutive pulse. This process is complex because the required pulse magnitude for linear formation depends on the EDL concentration after the previous pulse, the instantaneous formation and dissipation rates at that concentration, as well as the pulse width and frequency. Calculating all this information for each pulse in real-time is not feasible in practical applications. Therefore, we developed a predictive Linear Ionic Weight Update Solver (LIWUS) in Python. Details of the solver are discussed in the Methods section. To train the model, steady state step voltages ranging from 0.25 V to 3 V were first simulated in COMSOL and the corresponding EDL concentrations and rates extracted, as depicted in Fig. 5 (a) & (b). Note that this voltage range allows extrapolating the data to any voltage between 0 V and 3 V as depicted by the 3D plot in Fig. 5 (c), and the upper limit of 3 V was chosen to avoid electrochemical reactions which may change the EDLT channel conductance irreversibly.[28] Using the voltage and concentration-dependent rate data (both formation and dissipation) for training, LIWUS can predict the voltages that need to be applied to achieve a linear response. To ensure the predictive solver is applicable to different material systems and applications, the initial pulse magnitude, pulse width, frequency, number of pulses, and the slope of the desired linear increase (which relates to the on/off ratio) are all designed as variables in the model and can be defined by the user. Fig. 5 (d) illustrates the LIWUS prediction for 40 pulses, consisting of 20 potentiation pulses and 20 depression pulses, starting from a pulse magnitude of 0.4 V with a 1 ms pulse width at 500 Hz. Fig. 5 (f) shows the COMSOL simulation based on LIWUS predicted voltage pulse inputs, and the EDL reached a maximum concentration of 4.45 x $10^{13}$ ions/$cm^2$ and $\Delta n$ of 4.27 x $10^{13}$ ions/$cm^2$ for both potentiation and depression, with an $R^2$ of 0.99. As a comparison, fixed magnitude of input pulses with the same 500 Hz frequency resulted in a nonlinear change of conductance and a $\Delta n$ of 6.33 x $10^{13}$ ions/$cm^2$, as shown in Fig. 5 (e). Note that using a lower starting voltage (0.4 V) for the linear weight update compared to the nonlinear weight update (1 V) resulted in a lower $\Delta n$ for the former, but this range can be increased with a higher voltage starting pulse and/or with a greater number of pulses. A pulsing scheme predicted by LIWUS for 70 potentiation and depression pulses each with its corresponding COMSOL output is shown in Fig. S3, which resulted in a $\Delta n$ of 3.05 x $10^{14}$ ions/$cm^2$. Trained using pulsing-independent steady state data, LIWUS can predict the required pulse voltages for any number of pulses, pulse widths, frequencies, and linearities, making it suitable for various EDLT material systems and applications with differing ion mobility and channel conductivity. Although the simulation results in Fig. 5 were obtained with a 50 nm x 50 nm geometry, LIWUS can be trained and used to predict voltages for larger geometries or different gate-to-channel distances. For example, LIWUS predictions and simulation results for a 1 μm x 1 μm geometry with the same input parameters as Fig. 5 (d) are shown in Figure S3.

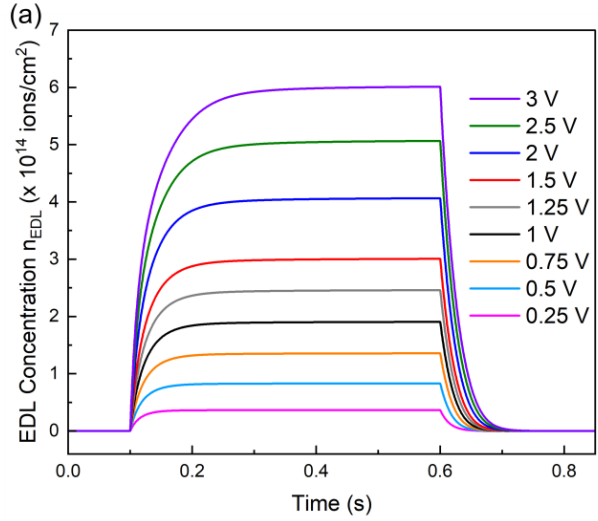
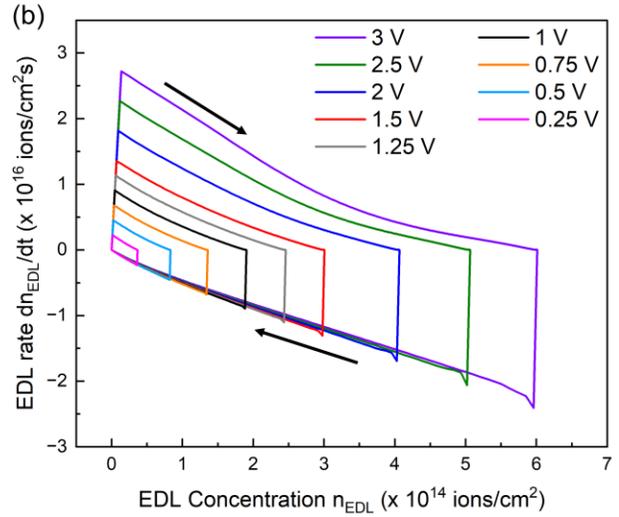
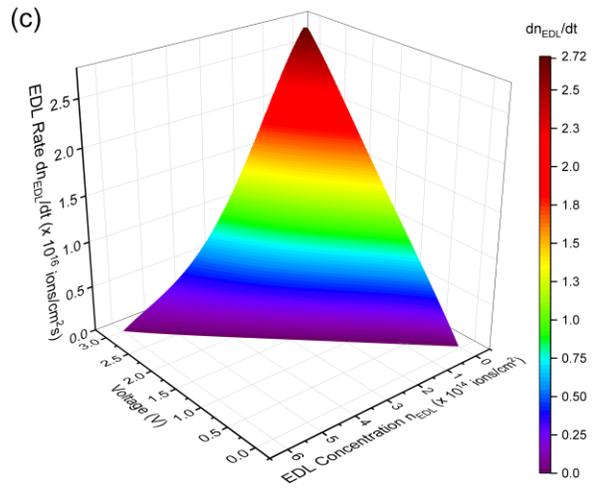
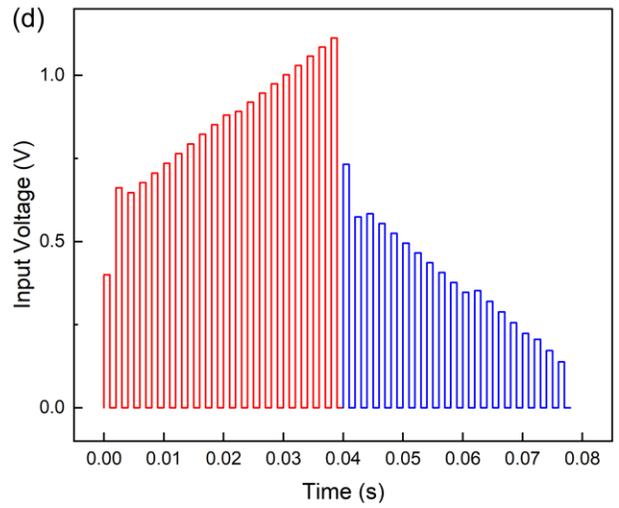
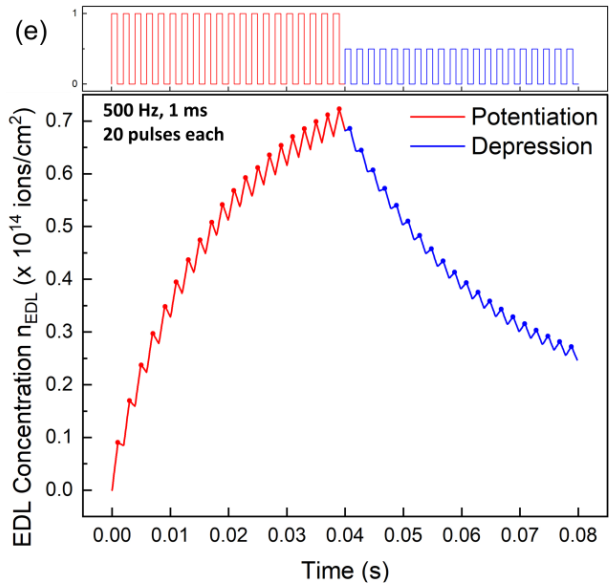
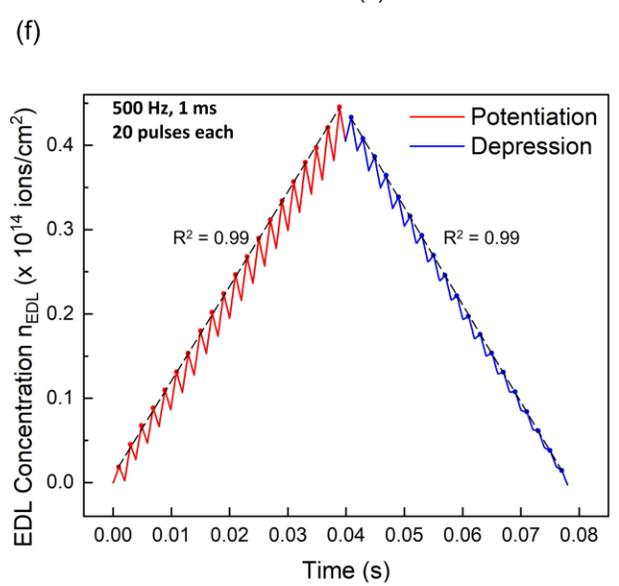

**Figure 5**: (a) Steady state EDL concentration simulated by step voltages from 0.25 V to 3 V, and (b) their corresponding rates as a function of concentration. (c) 3D plot of voltage & concentration-dependent EDL rates extrapolated from (b) and used as training data for LIWUS. (d) An example set of input voltages predicted by LIWUS to obtain linear plasticity, and (f) its corresponding response simulated by COMSOL. (e) Linear voltage pulses leading to nonlinear plasticity.

Now that linear synaptic weight updates can be produced with the help of LIWUS, the network-level performance of the devices needs to be evaluated to characterize the improvements of linear versus nonlinear plasticity. An artificial neural network based on a custom multi-layer perceptron (MLP) model was designed. The plasticity characteristics of the COMSOL-simulated devices (both nonlinear and linear, as shown in Fig. 5 (e) and 5 (f)) were fed into the neural network to perform supervised learning with image classification using the Modified National Institute of Standards and Technology (MNIST) dataset. Since the weight updates are achieved via pulse trains, the neural network model is amenable to future SNN applications. While the MNIST dataset is a widely used benchmark for image classification, the architecture of the artificial neural network could vary depending on the task, the nature of the input data, the availability of computational resources, and the type of learning.[61–63] The schematic of our MLP is illustrated in Fig. 6 (a) and can be extended to a crossbar array of synaptic weight layers containing EDLTs as shown in Fig. 6 (b). The MLP consists of 784 input neurons and 10 output nodes, corresponding to the 28 x 28-pixel MNIST image and the 10 digits respectively. The hidden layer contains 300 nodes, each one connected to each of the input and output nodes through synapses, the weights of which are derived from the device simulations in COMSOL. Details on neural network training are discussed in the Methods section.

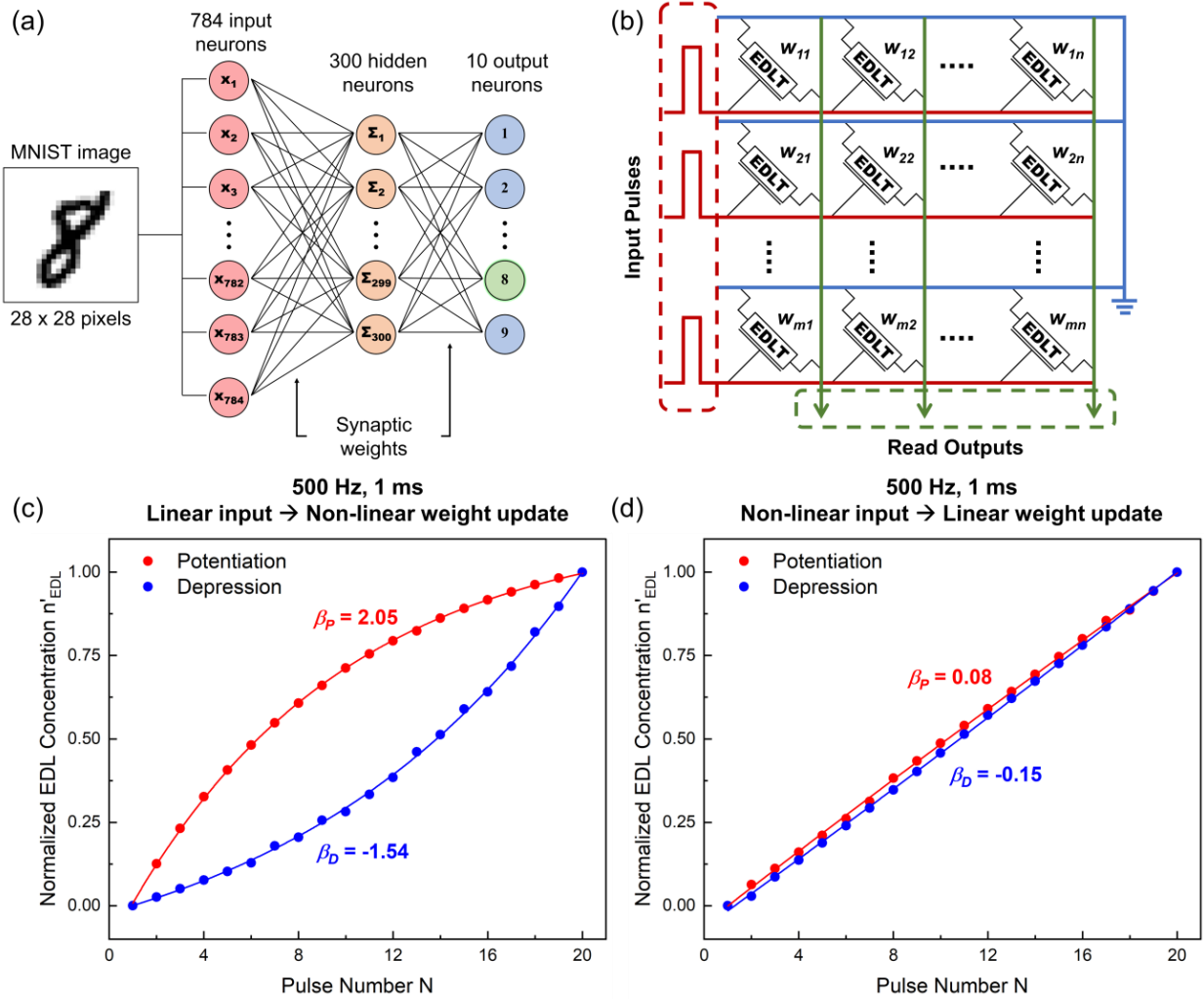

**Figure 6**: Schematics of (a) the three-layer perceptron and (b) the crossbar array of EDLTs. (c) & (d) Normalized EDL concentrations of nonlinear and linear weight updates as a function of pulse number N, respectively.

Previous studies on synaptic devices based on emerging materials have evaluated neural network performance by translating their devices' weight update characteristics to the network and training the network accordingly.[10,22,26,27,64–68] Building on those successful demonstrations, we designed and tested our neural network with the following focuses and additions: 1) incorporating an early stopping function, which is crucial for achieving optimal network performance while balancing the energy and time cost, and ensuring fair comparisons of different networks' performance; 2) benchmarking validation and testing accuracies in addition to training accuracies, as they better reflect the network's actual performance; and 3) performing 10 simulations at each testing parameter to track inconsistencies and quantify the variances produced by the network.

To extract information from COMSOL simulation for neural network training, EDL concentrations with nonlinear and linear weight updates were normalized in Fig. 6 (c) and (d), respectively. Note that the concentration states in COMSOL simulations correspond to the conductance states of a device in experimental settings.[29,34,50] The asymmetry and nonlinearity were extracted collectively as $\beta_P$ (for potentiation) & $\beta_D$ (for depression) from the normalized concentration ($n'_{EDL}$), using the following equation,

$$n'_{EDL} = \begin{cases} \frac{1-\exp(-\beta_P N)}{1-\exp(-\beta_P)} \\ \frac{\ln\{1-N[1-\exp(-\beta_D)]\}}{-\beta_D} \end{cases} \quad (1)$$

The asymmetric nonlinearity (ANL) parameters obtained were $\beta_P$ = 2.05 & 0.08 and $\beta_D$ = -1.54 & -0.15 for the nonlinear and linear responses, respectively. ANL parameters have significant impact on the accuracy of learning. For an ideal neural network with linear and symmetric weight update, this parameter approaches zero and does not require normalization,[29,69,70] which reduces processing steps and allows faster training of neural networks. The MNIST dataset consists of 60,000 training images and 10,000 testing images. 90% of the training images were used to train the model while the rest 10% was used as a validation set to monitor the training progress (i.e., 6000 images were used to test the ANN's performance after *each* training cycle). During training, the ANN adjusts its internal biases based on the patterns seen to learn the data's behavior and makes accurate predictions on unseen data. At the same time, the validation set helps prevent under- or overfitting to the training data. As mentioned above we adopted an early stopping function to monitor the validation accuracy, and training is stopped when the accuracy does not increase or decrease by the tolerance of 0.1% over 10 epochs, which is sufficient to observe any nontrivial changes to the model performance.[71–73] Although the number of training epochs required to obtain a good fit depends on the model architecture, the early stopping function allows comparing different models at their optimal performance. This optimally fit model is finally used to test performance of the neural network with the test dataset (10,000 unseen images). All the neural network simulations were conducted 10 times for each set of testing parameters to facilitate statistical analysis of the data and extract the error rate from the simulations.

Fig. 7 (a) illustrates how the recognition accuracies (validation and testing) change as the number of neurons in the hidden layer varies from 5 to 300. Fig. 7 (b) displays the accuracies following each training epoch, as well as the testing accuracy of the final (fully trained) model with 300 nodes in the hidden layer. The corresponding training accuracies are shown in Fig. S4. All data were averaged over ten simulations with the error bars indicating one standard deviation. The neural network with a linear weight update attains higher accuracies (validation accuracy of 98.24% and testing accuracy of 97.95%) compared to the nonlinear weight update (96.62% and 96.16%, respectively), as shown in Fig. 7 (b), confirming that a network is more effective with linear weight update. The early stopping function used with our network had *patience* set as 10 and *min_delta* of 0.001 (0.1%). Therefore, an absolute change in validation accuracy of less than 0.1% over 10 epochs acts as the stopping criteria for the model. These parameters were selected with the "cost-to-performance" in consideration as it is always a trade-off between the training time and error obtained. Longer training times imply lower loss or error but studies have shown only a marginal improvement in performance (~4%) over exponentially longer training times (4x).[71–74] With this implemented, the linear weight update model takes 5 fewer epochs (21 vs 26 epochs) to reach its optimal performance as shown in Fig. 7 (b), which can potentially translate to *up to* 19% less energy consumption for training. Recently, Artelnics (using similar hardware) reported their GPU energy consumption to be 4.5 kWh with TensorFlow.[75] If our ANN were to scale linearly, this would correspond to up to 0.86 kWh less energy consumed with the linear weight update model. However, more experimentation on larger networks would be required to generalize how energy consumption scales with size. Additionally, the linear weight update model shows less fluctuations in its prediction accuracy than the nonlinear weight update model (Fig. 7 (b)). The nonlinear weight update model shows a higher variance, with the standard deviation as high as 2.2% while the linear weight update model only deviates by a maximum of 0.3%. The nonlinear model converges at a lower accuracy while consuming more power (longer training time). Not surprisingly, the accuracy also increases with the increase in number of hidden neurons – for example, linear model's testing accuracy increased from 83.66% to 97.95% when the number of hidden neurons increased from 5 to 300.

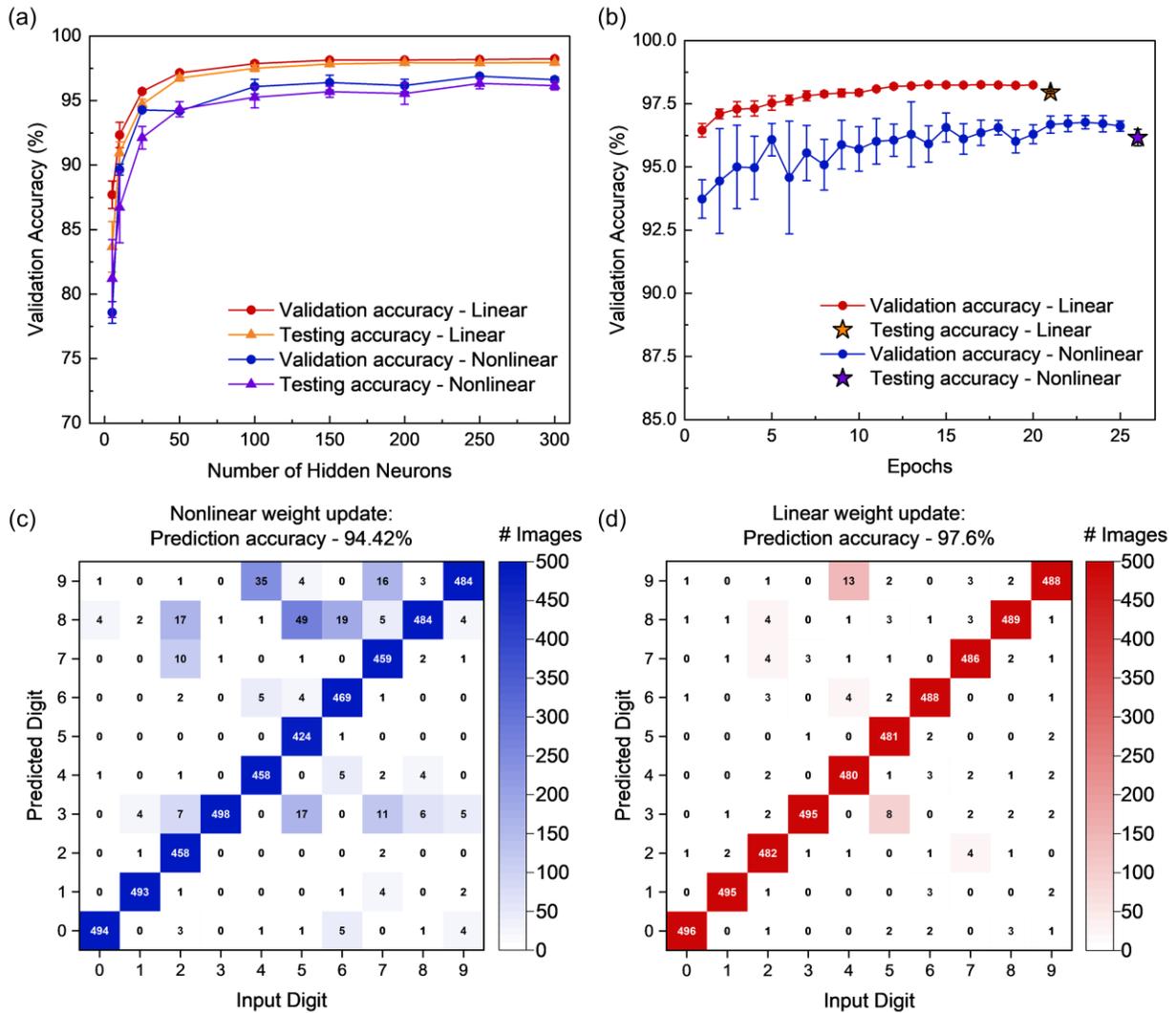

**Figure 7**: (a) Validation and testing accuracies vs. number of neurons in the hidden layer, (b) Validation accuracies extracted from the SNN vs. the number of training epochs, with the stars depicting testing accuracies of the final trained model, (c) & (d) confusion matrices showing the number of digits predicted correct and wrong by the final trained model against the actual input digit, for nonlinear and linear weight updates respectively.

After the models were fully trained, 5000 images (500 images for each digit) from the testing dataset were introduced with the task of classifying the images. The resulting confusion matrices are shown in Fig. 7 (c) & (d) for nonlinear and linear weight update models, respectively. The neural network with a nonlinear weight update resulted in a prediction accuracy of 94.42% (4721 images classified accurately); in comparison, the linear weight update model resulted in a 97.6% (4880 images) prediction accuracy. The confusion matrices after one training epoch are shown in Fig. S5, and the accuracy difference is larger (88% for nonlinear, 94% for linear). In summary, the linear weight update model shows 1.5 – 4.2% higher prediction (testing) accuracy, up to a 9.1% higher validation accuracy and a standard deviation less than a tenth than that of its nonlinear counterpart. The difference in prediction accuracies between the two models are statistically significant, with a $p = 7.28 \times 10^{-8} < 0.0001$ (paired t-test). We expect this difference to increase in the real-world scenario, when a more complex neural network with multiple hidden layers and more nodes are needed to perform more sophisticated tasks. For example, a memristor-based network with 2 hidden layers showed about 10% difference between a nonlinear, asymmetric model and an ideal, linear model.[76] Sun et al. showed over 12% difference in accuracies between an ideal, linear model and a symmetric, nonlinear model with the ANL parameters $β = ±1$, using the Canadian Institute For Advanced Research (CIFAR-10) dataset (images consisting of animals and automobiles).[77]

**CONCLUSIONS**

Finite element modeling was utilized to investigate the ion dynamics within an EDL capacitor by solving modified Poisson-Nernst-Planck (mPNP) equations. The simulation successfully captured the saturation behavior of the EDL ion concentration in response to voltage pulses of fixed magnitude, mirroring the nonlinear and saturating weight update observed experimentally in EDLT based synaptic devices. With 1 ms, 1 V pulses at 500 Hz, the ion concentration saturates at

~38% of the steady state concentration at 1 V. Analysis revealed both the EDL formation and dissipation rate to be dependent on the instantaneous EDL concentration. As the concentration increases, the formation rate decreases while dissipation rate increases, leading to saturation and nonlinear behavior observed in experimental studies.

Trained on steady state voltage and rate dependent concentration, a predictive model (LIWUS) was developed in Python which was able to predict a set of input voltages for user-defined parameters including pulse width, frequency, and initial pulse voltage, to achieve a linear weight update. COMSOL simulation with LIWUS predicted voltages resulted in a linear weight update in both 50 x 50 nm$^2$ and 1 x 1 μm$^2$ geometries, indicating that the predictive solver can potentially extend to a wide range of input parameters and device geometries. Lastly, a custom ANN was designed based on these modeled plasticities and tested for MNIST images classification. An early stopping function was implemented to achieve optimal network performance and ensure fair comparisons of different networks. A linear weight update-based model required 19% fewer (i.e. 5) epochs to reach optimal performance and exhibited a 1.5 – 4.2% improvement in recognition accuracies and less than a tenth of the standard deviation than that produced by the nonlinear model. These differences are statistically significant and are expected to increase for more sophisticated tasks requiring complex neural networks with multiple hidden layers and more nodes.

**METHODS**

**Finite element modeling and python predictive solver developing**

The behavior of the EDL capacitor was modeled using COMSOL Multiphysics, with the electrostatics module and a partial differential equation making the modified Poisson-Nernst-Planck (mPNP) equation

$$\frac{\partial c_\pm}{\partial t} = \nabla \cdot \left( D_\pm \nabla c_\pm + D_\pm \frac{e}{k_b T} z_\pm c_\pm \nabla V + \gamma \right)$$

where $c$ is the local ion concentration (ions/m$^3$), $D$ is the diffusivity (m$^2$/s), $k_bT$ is the thermal energy (J), $e$ is the elementary charge (C), $z$ is the ion charge number ($\pm 1$), and $V$ is electric potential (V) extracted from Poisson's equation,

$$\nabla^2(-\varepsilon V) = e(c_+ - c_-)$$

Here $\varepsilon$ is the permittivity of the electrolyte. The ion size is accounted for by the steric repulsion factor $\gamma$ and was derived by Kilic et al.[78] as

$$\gamma = D_\pm c_\pm a_\pm^3 \frac{\nabla(c_+ + c_-)}{1 - a_-^3 c_- - a_+^3 c_+}$$

where $a$ is the ion diameter. For a Li-like ion, diffusivity of 10$^{-12}$ cm$^2$/s and ion diameter of 0.2 nm were used.[52,79] A one ionic radius (0.1 nm) thick Stern layer was also established in the model, at each electrolyte-metal interface. The solid electrolyte capacitor has dimensions of 50 nm x 50 nm, and the metal contacts were modeled using boundary conditions where the top electrode was set to the applied bias and the bottom electrode was grounded (V = 0). Based on experimental values, the electrolyte permittivity value of 10 was chosen.[80] The time dependent solver used a time step size of 0.1 – 0.5 ms, depending on the pulse width. The solver used a triangular mesh that was made exponentially smaller and denser near the Stern layers, until ion concentrations were independent of mesh size, to more accurately model surface interactions and concentrations while keeping the computation time minimal.[81]

The Linear Ionic Weight Update Solver (LIWUS) was implemented in Python, leveraging the Matplotlib and NumPy libraries. The solver was trained on steady state data spanning a voltage range of 0.25 to 3 V within the electrochemical window of the electrolyte. This dataset was fitted to equations that enable accurate prediction of ion concentration changes for any concentration-voltage pair within this range. LIWUS allows the following input parameters: frequency, initial input magnitude, rate of change of concentration (slope for linearity), time step size, pulse duration, and starting concentration. The solver simulates the first pulse based on user-defined inputs for the initial period of the specified frequency. Subsequently, a recursive algorithm, combined with the desired rate of concentration change, calculates the optimal voltages necessary to maintain a linear ionic weight update throughout the specified number of pulses. The fitting process was conducted on a 50 nm x 50 nm device, but LIWUS can be potentially extended to any device geometry with re-training. LIWUS predictions and simulation results for a 1 μm x 1 μm geometry are shown in Figure S3.

**Artificial neural network**

The ANN developed for this study was implemented with the open-source software library TensorFlow and its interface, Keras in Python. The model was trained on the Modified National Institute of Standards and Technology (MNIST) handwritten digits dataset, a set of 60,000 numbers. The ANN's accuracy extracted is the ability of the network to correctly assess the digit in each image over the entire dataset. 6,000 images (10%) from the training set were reserved for validation testing in all the simulations. The model consists of three layers: an input layer of 784 nodes (one node assigned to each pixel in the 28 x 28-pixel images), a hidden (dense) layer of 300 nodes, and a final output layer of 10 nodes, each of which corresponds to a digit from 0 – 9. For the simulations involving different network sizes (Fig. 7 (a) & Fig. S4), the number of neurons in the hidden layer were selected as 5, 10, 25, 50, 100, 150, 200, 250, and 300. The model employed a Rectified Linear Unit (ReLU) activation function and was trained with the Stochastic Gradient Descent (SGD) function. A custom Keras layer class was used to quantize each weight to 20 discrete values. Weight updates based on the ANL parameters were implemented to accurately model the EDL device's behavior. Accuracies extracted from 10 simulations for the same model parameters were averaged to obtain statistical results. The number of epochs each model was trained to was determined dynamically with Keras' native early stopping function, which automatically stops the network's training after a user-defined number of epochs if no improvement is observed with the validation accuracy. This number acts as the algorithm's "patience" value and was set to 10, with a *min_delta* (tolerance) of 0.001 (0.1%), which is sufficient to obtain a fully trained and optimized model. 5000 images from the testing dataset were used for the confusion matrices. Since all the digits are not equally distributed over the entire testing dataset, additional code was written in Python to extract 500 images of each digit and have the order shuffled. This final dataset was then used to evaluate the classification accuracies of the fully trained model and the classified number of digits were extracted as a confusion matrix.

### STATISTICAL ANALYSIS

Paired t-test was performed according to IBM SPSS Statistics 25.0.[82]

### ACKNOWLEDGMENTS

This work was partially funded by RIT College of Science Dean's Research Initiation Grant. The authors thank Claire Dantzlerward at Rochester Institute of Technology for the schematic of

**REFERENCES**

(1) Zou, X.; Xu, S.; Chen, X.; Yan, L.; Han, Y. Breaking the von Neumann Bottleneck: Architecture-Level Processing-in-Memory Technology. *Science China Information Sciences* **2021**, *64* (6), 160404.

(2) Hennessy, J. L.; Patterson, D. A. *Computer Architecture: A Quantitative Approach*, 6th ed.; Elsevier: New York, 2018.

(3) Kim, Y.; Kiemb, M.; Park, C.; Jung, J.; Choi, K. Resource Sharing and Pipelining in Coarse-Grained Reconfigurable Architecture for Domain-Specific Optimization. **2005**.

(4) Jouppi, N. P.; Young, C.; Patil, N.; Patterson, D. A Domain-Specific Architecture for Deep Neural Networks. *Commun ACM* **2018**, *61* (9), 50–59. https://doi.org/10.1145/3154484.

(5) Muralidhar, R.; Borovica-Gajic, R.; Buyya, R. Energy Efficient Computing Systems: Architectures, Abstractions and Modeling to Techniques and Standards. *ACM Computing Surveys (CSUR)* **2022**, *54* (11s), 1–37.

(6) Mead, C. Neuromorphic Electronic Systems. *Proceedings of the IEEE* **1990**, *78* (10), 1629–1636.

(7) Worden, K.; Tsialiamanis, G.; Cross, E. J.; Rogers T J. Artificial Neural Networks. In *Machine Learning in Modeling and Simulation: Methods and Applications*; Rabczuk, T., Bathe, K.-J., Eds.; Springer: Cambridge, 2023; pp 85–119.

(8) Schuman, C. D.; Kulkarni, S. R.; Parsa, M.; Mitchell, J. P.; Date, P.; Kay, B. Opportunities for Neuromorphic Computing Algorithms and Applications. *Nature Computational Science*. Springer Nature January 1, 2022, pp 10–19. https://doi.org/10.1038/s43588-021-00184-y.

(9) Cho, H.; Lee, D.; Ko, K.; Lin, D. Y.; Lee, H.; Park, S.; Park, B.; Jang, B. C.; Lim, D. H.; Suh, J. Double-Floating-Gate van Der Waals Transistor for High-Precision Synaptic Operations. *ACS Nano* **2023**, *17* (8), 7384–7393. https://doi.org/10.1021/acsnano.2c11538.

(10) Seo, S.; Kim, B.; Kim, D.; Park, S.; Kim, T. R.; Park, J.; Jeong, H.; Park, S. O.; Park, T.; Shin, H.; Kim, M. S.; Choi, Y. K.; Choi, S. The Gate Injection-Based Field-Effect Synapse Transistor with Linear Conductance Update for Online Training. *Nat Commun* **2022**, *13* (1). https://doi.org/10.1038/s41467-022-34178-9.

(11) Furber, S. B.; Lester, D. R.; Plana, L. A.; Garside, J. D.; Painkras, E.; Temple, S.; Brown, A. D. Overview of the SpiNNaker System Architecture. *IEEE transactions on computers* **2012**, *62* (12), 2454–2467.

(12) Davies, M.; Wild, A.; Orchard, G.; Sandamirskaya, Y.; Guerra, G. A. F.; Joshi, P.; Plank, P.; Risbud, S. R. Advancing Neuromorphic Computing with Loihi: A Survey of Results and Outlook. *Proceedings of the IEEE* **2021**, *109* (5), 911–934. https://doi.org/10.1109/JPROC.2021.3067593.

(13) Roy, K.; Jaiswal, A.; Panda, P. Towards Spike-Based Machine Intelligence with Neuromorphic Computing. *Nature* **2019**, *575* (7784), 607–617. https://doi.org/10.1038/s41586-019-1677-2.

(14) Mead, C. How We Created Neuromorphic Engineering. *Nature Electronics*. Nature Research July 1, 2020, pp 434–435. https://doi.org/10.1038/s41928-020-0448-2.

(15) Patton, R.; Schuman, C.; Kulkarni, S.; Parsa, M.; Mitchell, J. P.; Haas, N. Q.; Stahl, C.; Paulissen, S.; Date, P.; Potok, T.; Sneider, S. Neuromorphic Computing for Autonomous Racing. In *ACM International Conference Proceeding Series*; Association for Computing Machinery, 2021. https://doi.org/10.1145/3477145.3477170.

(16) Caire, M. J.; Reddy, V.; Varacallo, M. *Physiology, Synapse*; 2018.

(17) Gabbiani, F.; Cox, S. Chapter 12. Synaptic Transmission and Quantal Release. Mathematics for Neuroscientists. 1st ed. Academic Press 2010.

(18) Bean, B. P. The Action Potential in Mammalian Central Neurons. *Nat Rev Neurosci* **2007**, *8* (6), 451–465.

(19) Purves, D.; Augustine, G. J.; Fitzpatrick, D.; Katz, L. C.; LaMantia, A. S.; McNamara, J. O.; Williams, S. M. The Organization of the Nervous System. *Neuroscience. Purves D, Augustine GJ, Fitzpatrick D, Katz LC, LaMantia AS, McNamara JO, et al, editors. Sunderland, MA: Sinauer Associates* **2001**.

(20) Markram, H.; Gerstner, W.; Sjöström, P. J. Spike-Timing-Dependent Plasticity: A Comprehensive Overview. *Front Synaptic Neurosci* **2012**, *4*, 2.

(21) Caporale, N.; Dan, Y. Spike Timing–Dependent Plasticity: A Hebbian Learning Rule. *Annu. Rev. Neurosci.* **2008**, *31*, 25–46.

(22) Moon, K.; Lim, S.; Park, J.; Sung, C.; Oh, S.; Woo, J.; Lee, J.; Hwang, H. RRAM-Based Synapse Devices for Neuromorphic Systems. *Faraday Discuss* **2019**, *213*, 421–451. https://doi.org/10.1039/c8fd00127h.

(23) Shen, Z.; Zhao, C.; Qi, Y.; Xu, W.; Liu, Y.; Mitrovic, I. Z.; Yang, L.; Zhao, C. Advances of RRAM Devices: Resistive Switching Mechanisms, Materials and Bionic Synaptic Application. *Nanomaterials*. MDPI AG August 1, 2020, pp 1–31. https://doi.org/10.3390/nano10081437.

(24) Yang, J. Q.; Wang, R.; Wang, Z. P.; Ma, Q. Y.; Mao, J. Y.; Ren, Y.; Yang, X.; Zhou, Y.; Han, S. T. Leaky Integrate-and-Fire Neurons Based on Perovskite Memristor for Spiking Neural Networks. *Nano Energy* **2020**, *74*. https://doi.org/10.1016/j.nanoen.2020.104828.

(25) Nandakumar, S. R.; Boybat, I.; Le Gallo, M.; Eleftheriou, E.; Sebastian, A.; Rajendran, B. Experimental Demonstration of Supervised Learning in Spiking Neural Networks with Phase-Change Memory Synapses. *Sci Rep* **2020**, *10* (1). https://doi.org/10.1038/s41598-020-64878-5.

(26) Yu, S.; Hur, J.; Luo, Y. C.; Shim, W.; Choe, G.; Wang, P. Ferroelectric HfO2-Based Synaptic Devices: Recent Trends and Prospects. *Semicond Sci Technol* **2021**, *36* (10). https://doi.org/10.1088/1361-6641/ac1b11.

(27) Zheng, C.; Liao, Y.; Wang, J.; Zhou, Y.; Han, S. T. Flexible Floating-Gate Electric-Double-Layer Organic Transistor for Neuromorphic Computing. *ACS Appl Mater Interfaces* **2022**, *14* (51), 57102–57112. https://doi.org/10.1021/acsami.2c20925.

(28) Xu, K.; Fullerton-Shirey, S. K. Electric-Double-Layer-Gated Transistors Based on Two-Dimensional Crystals: Recent Approaches and Advances. *JPhys Materials*. IOP Publishing Ltd July 1, 2020. https://doi.org/10.1088/2515-7639/ab8270.

(29) Yang, C.; Shang, D.; Liu, N.; Fuller, E. J.; Agrawal, S.; Talin, A. A.; Li, Y.; Shen, B.; Sun, Y. All-solid-state Synaptic Transistor with Ultralow Conductance for Neuromorphic Computing. *Adv Funct Mater* **2018**, *28* (42), 1804170.

(30) Wu, G.; Feng, P.; Wan, X.; Zhu, L.; Shi, Y.; Wan, Q. Artificial Synaptic Devices Based on Natural Chicken Albumen Coupled Electric-Double-Layer Transistors. *Sci Rep* **2016**, *6*. https://doi.org/10.1038/srep23578.

(31) Wan, C. J.; Liu, Y. H.; Zhu, L. Q.; Feng, P.; Shi, Y.; Wan, Q. Short-Term Synaptic Plasticity Regulation in Solution-Gated Indium-Gallium-Zinc-Oxide Electric-Double-Layer Transistors. *ACS Appl Mater Interfaces* **2016**, *8* (15), 9762–9768. https://doi.org/10.1021/acsami.5b12726.

(32) Wen, J.; Zhu, L. Q.; Fu, Y. M.; Xiao, H.; Guo, L. Q.; Wan, Q. Activity Dependent Synaptic Plasticity Mimicked on Indium-Tin-Oxide Electric-Double-Layer Transistor. *ACS Appl Mater Interfaces* **2017**, *9* (42), 37064–37069. https://doi.org/10.1021/acsami.7b13215.

(33) Zhou, J.; Liu, N.; Zhu, L.; Shi, Y.; Wan, Q. Energy-Efficient Artificial Synapses Based on Flexible IGZO Electric-Double-Layer Transistors. *IEEE Electron Device Letters* **2015**, *36* (2), 198–200. https://doi.org/10.1109/LED.2014.2381631.

(34) He, Y.; Yang, Y.; Nie, S.; Liu, R.; Wan, Q. Electric-Double-Layer Transistors for Synaptic Devices and Neuromorphic Systems. *Journal of Materials Chemistry C*. Royal Society of Chemistry 2018, pp 5336–5352. https://doi.org/10.1039/c8tc00530c.

(35) Peng, C.; Jiang, W.; Li, Y.; Li, X.; Zhang, J. Photoelectric IGZO Electric-Double-Layer Transparent Artificial Synapses for Emotional State Simulation. *ACS Appl Electron Mater* **2019**, *1* (11), 2406–2414.

(36) Jiang, J.; Hu, W.; Xie, D.; Yang, J.; He, J.; Gao, Y.; Wan, Q. 2D Electric-Double-Layer Phototransistor for Photoelectronic and Spatiotemporal Hybrid Neuromorphic Integration. *Nanoscale* **2019**, *11* (3), 1360–1369.

(37) Yuan, H.; Shimotani, H.; Tsukazaki, A.; Ohtomo, A.; Kawasaki, M.; Iwasa, Y. High-Density Carrier Accumulation in ZnO Field-Effect Transistors Gated by Electric Double Layers of Ionic Liquids. *Adv Funct Mater* **2009**, *19* (7), 1046–1053. https://doi.org/10.1002/adfm.200801633.

(38) Awate, S. S.; Mostek, B.; Kumari, S.; Dong, C.; Robinson, J. A.; Xu, K.; Fullerton-Shirey, S. K. Impact of Large Gate Voltages and Ultrathin Polymer Electrolytes on Carrier Density in Electric-Double-Layer-Gated Two-Dimensional Crystal Transistors. *ACS Appl Mater Interfaces* **2023**, *15* (12), 15785–15796.

(39) Xu, H.; Fathipour, S.; Kinder, E. W.; Seabaugh, A. C.; Fullerton-Shirey, S. K. Reconfigurable Ion Gating of 2H-MoTe2 Field-Effect Transistors Using Poly (Ethylene Oxide)-CsClO4 Solid Polymer Electrolyte. *ACS Nano* **2015**, *9* (5), 4900–4910.

(40) Kim, S. H.; Hong, K.; Xie, W.; Lee, K. H.; Zhang, S.; Lodge, T. P.; Frisbie, C. D. Electrolyte-gated Transistors for Organic and Printed Electronics. *Advanced Materials* **2013**, *25* (13), 1822–1846.

(41) Zhu, J.; Yang, Y.; Jia, R.; Liang, Z.; Zhu, W.; Rehman, Z. U.; Bao, L.; Zhang, X.; Cai, Y.; Song, L. Ion Gated Synaptic Transistors Based on 2D van Der Waals Crystals with Tunable Diffusive Dynamics. *Advanced Materials* **2018**, *30* (21), 1800195.

(42) Xu, K.; Islam, M. M.; Guzman, D.; Seabaugh, A. C.; Strachan, A.; Fullerton-Shirey, S. K. Pulse Dynamics of Electric Double Layer Formation on All-Solid-State Graphene Field-Effect Transistors. *ACS Appl Mater Interfaces* **2018**, *10* (49), 43166–43176.

(43) Min, S. Y.; Cho, W. J. CMOS-Compatible Synaptic Transistor Gated by Chitosan Electrolyte-Ta2O5 Hybrid Electric Double Layer. *Sci Rep* **2020**, *10* (1). https://doi.org/10.1038/s41598-020-72684-2.

(44) Agarwal, S.; Plimpton, S. J.; Hughart, D. R.; Hsia, A. H.; Richter, I.; Cox, J. A.; James, C. D.; Marinella, M. J. Resistive Memory Device Requirements for a Neural Algorithm Accelerator. In *2016 International Joint Conference on Neural Networks (IJCNN)*; IEEE, 2016; pp 929–938.

(45) Kim, T.; Hu, S.; Kim, J.; Kwak, J. Y.; Park, J.; Lee, S.; Kim, I.; Park, J.-K.; Jeong, Y. Spiking Neural Network (Snn) with Memristor Synapses Having Non-Linear Weight Update. *Front Comput Neurosci* **2021**, *15*, 646125.

(46) Wang, Y.; Li, F.; Sun, H.; Li, W.; Zhong, C.; Wu, X.; Wang, H.; Wang, P. Improvement of MNIST Image Recognition Based on CNN. In *IOP Conference Series: Earth and Environmental Science*; IOP Publishing, 2020; Vol. 428, p 12097.

(47) Zhou, J.; Liu, N.; Zhu, L.; Shi, Y.; Wan, Q. Energy-Efficient Artificial Synapses Based on Flexible IGZO Electric-Double-Layer Transistors. *IEEE Electron Device Letters* **2014**, *36* (2), 198–200.

(48) Kim, H.-S.; Park, H.; Cho, W.-J. Biocompatible Casein Electrolyte-Based Electric-Double-Layer for Artificial Synaptic Transistors. *Nanomaterials* **2022**, *12* (15), 2596.

(49) Lee, D.-H.; Park, H.; Cho, W.-J. Synaptic Transistors Based on PVA: Chitosan Biopolymer Blended Electric-Double-Layer with High Ionic Conductivity. *Polymers (Basel)* **2023**, *15* (4), 896.

(50) Feng, X.; Qiao, L.; Huang, J.; Ning, J.; Wang, D.; Zhang, J.; Hao, Y. A Novel CVD Graphene-Based Synaptic Transistors with Ionic Liquid Gate. *Nanotechnology* **2023**, *34* (21). https://doi.org/10.1088/1361-6528/acbc82.

(51) Wang, D.; Zhao, S.; Yin, R.; Li, L.; Lou, Z.; Shen, G. Recent Advanced Applications of Ion-Gel in Ionic-Gated Transistor. *npj Flexible Electronics*. Nature Research December 1, 2021. https://doi.org/10.1038/s41528-021-00110-2.

(52) Woeppel, A.; Xu, K.; Kozhakhmetov, A.; Awate, S.; Robinson, J. A.; Fullerton-Shirey, S. K. Single- versus Dual-Ion Conductors for Electric Double Layer Gating: Finite Element Modeling and Hall-Effect Measurements. *ACS Appl Mater Interfaces* **2020**, *12* (36), 40850–40858. https://doi.org/10.1021/acsami.0c08653.

(53) Zhang, K.; Jia, X.; Cao, K.; Wang, J.; Zhang, Y.; Lin, K.; Chen, L.; Feng, X.; Zheng, Z.; Zhang, Z.; Zhang, Y.; Zhao, W. High On/Off Ratio Spintronic Multi-Level Memory Unit for Deep Neural Network. *Advanced Science* **2022**, *9* (13). https://doi.org/10.1002/advs.202103357.

(54) Safayenikoo, P.; Akturk, I. Weight Update Skipping: Reducing Training Time for Artificial Neural Networks. *IEEE J Emerg Sel Top Circuits Syst* **2021**, *11* (4), 563–574. https://doi.org/10.1109/JETCAS.2021.3127907.

(55) Li, H. M.; Xu, K.; Bourdon, B.; Lu, H.; Lin, Y. C.; Robinson, J. A.; Seabaugh, A. C.; Fullerton-Shirey, S. K. Electric Double Layer Dynamics in Poly(Ethylene Oxide) LiClO4 on Graphene Transistors. *Journal of Physical Chemistry C* **2017**, *121* (31), 16996–17004. https://doi.org/10.1021/acs.jpcc.7b04788.

(56) Du, B. X.; Jiang, J. P.; Zhang, J. G.; Liu, D. S. Dynamic Behavior of Surface Charge on Double-Layer Oil-Paper Insulation under Pulse Voltage. *IEEE Transactions on Dielectrics and Electrical Insulation* **2016**, *23* (5), 2712–2719. https://doi.org/10.1109/TDEI.2016.005321.

(57) Kim, Y.; Kim, J.; Kong, J. J.; K Vijaya Kumar, B. V; Li, X. Verify Level Control Criteria for Multi-Level Cell Flash Memories and Their Applications. *EURASIP J Adv Signal Process* **2012**, *2012*, 1–13.

(58) Du, Z.; Li, S.; Wang, Y.; Fu, X.; Liu, F.; Wang, Q.; Huo, Z. Adaptive Pulse Programming Scheme for Improving the V Th Distribution and Program Performance in 3D NAND Flash Memory. *IEEE Journal of the Electron Devices Society* **2020**, *9*, 102–107.

(59) Nam, K.; Park, C.; Yoon, J.-S.; Jang, H.; Park, M. S.; Sim, J.; Baek, R.-H. Origin of Incremental Step Pulse Programming (ISPP) Slope Degradation in Charge Trap Nitride Based Multi-Layer 3D NAND Flash. *Solid State Electron* **2021**, *175*, 107930.

(60) Song, M. K.; Kang, J. H.; Zhang, X.; Ji, W.; Ascoli, A.; Messaris, I.; Demirkol, A. S.; Dong, B.; Aggarwal, S.; Wan, W.; Hong, S. M.; Cardwell, S. G.; Boybat, I.; Seo, J. S.; Lee, J. S.; Lanza, M.; Yeon, H.; Onen, M.; Li, J.; Yildiz, B.; del Alamo, J. A.; Kim, S.; Choi, S.; Milano, G.; Ricciardi, C.; Alff, L.; Chai, Y.; Wang, Z.; Bhaskaran, H.; Hersam, M. C.; Strukov, D.; Wong, H. S. P.; Valov, I.; Gao, B.; Wu, H.; Tetzlaff, R.; Sebastian, A.; Lu, W.; Chua, L.; Yang, J. J.; Kim, J. Recent Advances and Future Prospects for Memristive Materials, Devices, and Systems. *ACS Nano*. American Chemical Society July 11, 2023, pp 11994–12039. https://doi.org/10.1021/acsnano.3c03505.

(61) Beohar, D.; Rasool, A. Handwritten Digit Recognition of MNIST Dataset Using Deep Learning State-of-the-Art Artificial Neural Network (ANN) and Convolutional Neural Network (CNN). In *2021 International Conference on Emerging Smart Computing and Informatics, ESCI 2021*; Institute of Electrical and Electronics Engineers Inc., 2021; pp 542–548. https://doi.org/10.1109/ESCI50559.2021.9396870.

(62) Elizabeth Rani, G.; Sakthimohan, M.; Abhigna Reddy, G.; Selvalakshmi, D.; Keerthi, T.; Raja Sekar, R. MNIST Handwritten Digit Recognition Using Machine Learning. In *2022 2nd International Conference on Advance Computing and Innovative Technologies in Engineering, ICACITE 2022*; Institute of Electrical and Electronics Engineers Inc., 2022; pp 768–772. https://doi.org/10.1109/ICACITE53722.2022.9823806.

(63) Baldominos, A.; Saez, Y.; Isasi, P. A Survey of Handwritten Character Recognition with MNIST and EMNIST. *Applied Sciences (Switzerland)*. MDPI AG August 1, 2019. https://doi.org/10.3390/app9153169.

(64) Xiao, T. P.; Bennett, C. H.; Feinberg, B.; Agarwal, S.; Marinella, M. J. Analog Architectures for Neural Network Acceleration Based on Non-Volatile Memory. *Applied Physics Reviews*. American Institute of Physics Inc. September 1, 2020. https://doi.org/10.1063/1.5143815.

(65) Zivasatienraj, B.; Doolittle, W. A. Dynamical Memristive Neural Networks and Associative Self-Learning Architectures Using Biomimetic Devices. *Front Neurosci* 2023, *17*. https://doi.org/10.3389/fnins.2023.1153183.

(66) Das, H.; Schuman, C.; Chakraborty, N. N.; Rose, G. S. Enhanced Read Resolution in Reconfigurable Memristive Synapses for Spiking Neural Networks. *Sci Rep* 2024, *14* (1). https://doi.org/10.1038/s41598-024-58947-2.

(67) Luo, Y.; Yu, S. Accelerating Deep Neural Network In-Situ Training with Non-Volatile and Volatile Memory Based Hybrid Precision Synapses. *IEEE Transactions on Computers* 2020, *69* (8), 1113–1127. https://doi.org/10.1109/TC.2020.3000218.

(68) Zhang, H. T.; Park, T. J.; Zaluzhnyy, I. A.; Wang, Q.; Wadekar, S. N.; Manna, S.; Andrawis, R.; Sprau, P. O.; Sun, Y.; Zhang, Z.; Huang, C.; Zhou, H.; Zhang, Z.; Narayanan, B.; Srinivasan, G.; Hua, N.; Nazaretski, E.; Huang, X.; Yan, H.; Ge, M.; Chu, Y. S.; Cherukara, M. J.; Holt, M. V.; Krishnamurthy, M.; Shpyrko, O. G.; Sankaranarayanan, S. K. R. S.; Frano, A.; Roy, K.; Ramanathan, S. Perovskite Neural Trees. *Nat Commun* 2020, *11* (1). https://doi.org/10.1038/s41467-020-16105-y.

(69) Kayed, M.; Anter, A.; Mohamed, H. Classification of Garments from Fashion MNIST Dataset Using CNN LeNet-5 Architecture. In *2020 international conference on innovative trends in communication and computer engineering (ITCE)*; IEEE, 2020; pp 238–243.

(70) Kadam, S. S.; Adamuthe, A. C.; Patil, A. B. CNN Model for Image Classification on MNIST and Fashion-MNIST Dataset. *Journal of scientific research* 2020, *64* (2), 374–384.

(71) Waheed, A.; Goyal, M.; Gupta, D.; Khanna, A.; Hassanien, A. E.; Pandey, H. M. An Optimized Dense Convolutional Neural Network Model for Disease Recognition and Classification in Corn Leaf. *Comput Electron Agric* 2020, *175*. https://doi.org/10.1016/j.compag.2020.105456.

(72) Ashqar, B. A. M.; Abu-Naser, S. S. *Identifying Images of Invasive Hydrangea Using Pre-Trained Deep Convolutional Neural Networks*. www.ijeais.org/ijaer.

(73) Ammous, D.; Chabbouh, A.; Edhib, A.; Chaari, A.; Kammoun, F.; Masmoudi, N. Designing an Efficient System for Emotion Recognition Using CNN. *Journal of Electrical and Computer Engineering* 2023, *2023*. https://doi.org/10.1155/2023/9351345.

(74) Prechelt, L. *2 Early Stopping-But When?* http://www.ipd.ira.uka.de/~prechelt/.

(75) Rodrigo Ingelmo Vicente. *Energy consumption of TensorFlow and Neural Designer*. https://www.neuraldesigner.com/blog/energy-consumption-comparison/.

(76) Park, S. O.; Park, T.; Jeong, H.; Hong, S.; Seo, S.; Kwon, Y.; Lee, J.; Choi, S. Linear Conductance Update Improvement of CMOS-Compatible Second-Order Memristors for Fast and Energy-Efficient Training of a Neural Network Using a Memristor Crossbar Array. *Nanoscale Horiz* 2023, *8* (10), 1366–1376. https://doi.org/10.1039/d3nh00121k.

(77) Sun, X.; Yu, S. Impact of Non-Ideal Characteristics of Resistive Synaptic Devices on Implementing Convolutional Neural Networks. In *IEEE Journal on Emerging and Selected Topics in Circuits and Systems*; Institute of Electrical and Electronics Engineers Inc., 2019; Vol. 9, pp 570–579. https://doi.org/10.1109/JETCAS.2019.2933148.

(78) Kilic, M. S.; Bazant, M. Z.; Ajdari, A. Steric Effects in the Dynamics of Electrolytes at Large Applied Voltages. II. Modified Poisson-Nernst-Planck Equations. *Phys Rev E* **2007**, *75* (2), 21503.

(79) Peng, L.; Zhu, Y.; Chen, D.; Ruoff, R. S.; Yu, G. Two-Dimensional Materials for Beyond-Lithium-Ion Batteries. *Advanced Energy Materials*. Wiley-VCH Verlag June 8, 2016. https://doi.org/10.1002/aenm.201600025.

(80) Gonzalez-Serrano, K. A.; Seabaugh, A. C. Electrical Properties of 6 Nm to 19 Nm Thick Polyethylene Oxide Capacitors for Ion/Electron Functional Devices. *J Electron Mater* **2021**, *50* (6), 2956–2963. https://doi.org/10.1007/s11664-020-08716-4.

(81) Xu, K.; Liang, J.; Woeppel, A.; Bostian, M. E.; Ding, H.; Chao, Z.; McKone, J. R.; Beckman, E. J.; Fullerton-Shirey, S. K. Electric Double-Layer Gating of Two-Dimensional Field-Effect Transistors Using a Single-Ion Conductor. *ACS Appl Mater Interfaces* **2019**, *11* (39), 35879–35887.

(82) Mehta, C. R.; Patel, N. R. *IBM SPSS Exact Tests*.

Supporting Information

# Realizing Linear Synaptic Plasticity in Electric Double Layer-Gated Transistors for Improved Predictive Accuracy and Efficiency in Neuromorphic Computing

Nithil Harris Manimaran[1], Cori Sutton[2], Jake Streamer[3], Cory Merkel[4], and Ke Xu[1,2,5]*

[1]Microsystems Engineering, Rochester Institute of Technology, Rochester, New York 14623, United States

[2]School of Physics and Astronomy, Rochester Institute of Technology, Rochester, New York 14623, United States

[3]Multidisciplinary Study, Rochester Institute of Technology, Rochester, New York 14623, United States

[4]Department of Electrical and Computer Engineering, Rochester Institute of Technology, Rochester, New York 14623, United States

[5]School of Chemistry and Material Science, Rochester Institute of Technology, Rochester, New York 14623, United States

*Corresponding author – Email: ke.xu@rit.edu

# Supporting Information

## S1 – Depression with 5 ms pulses depicting nonlinearity

Depression with 5 ms pulses with frequencies ranging from 10 Hz to 100 Hz also shows saturation after a few pulses, similar to what was observed with potentiation (Fig. 3). With the 10 Hz pulses, no EDL accumulation was observed as the ions relax back to their resting state between subsequent pulses.

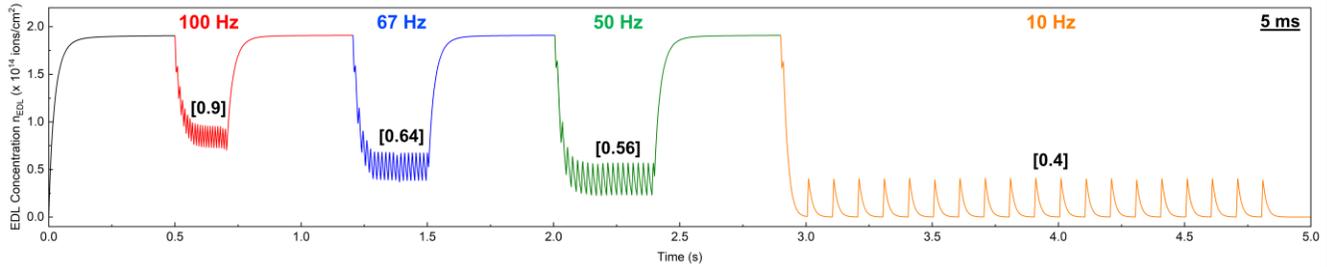

*Figure S1: EDL concentrations extracted from COMSOL as a response to 5 ms pulses with frequencies 100 Hz, 67 Hz, 50 Hz, & 10 Hz, illustrating nonlinearity with depression.*

## S2 – Effect of frequency on saturation EDL concentration

At a particular pulse width, the saturation EDL concentration ($n_{EDL}^{P}$) increases with frequency as the time between pulses grows shorter. Likewise, at a specific frequency, $n_{EDL}^{P}$ increases with pulse width as the on time increases. In addition, both the number of distinct states and usable concentration range $\Delta n$ increases with frequency and pulse width. Steady state EDL concentration ($n_{EDL}^{SS}$) is also indicated for reference.

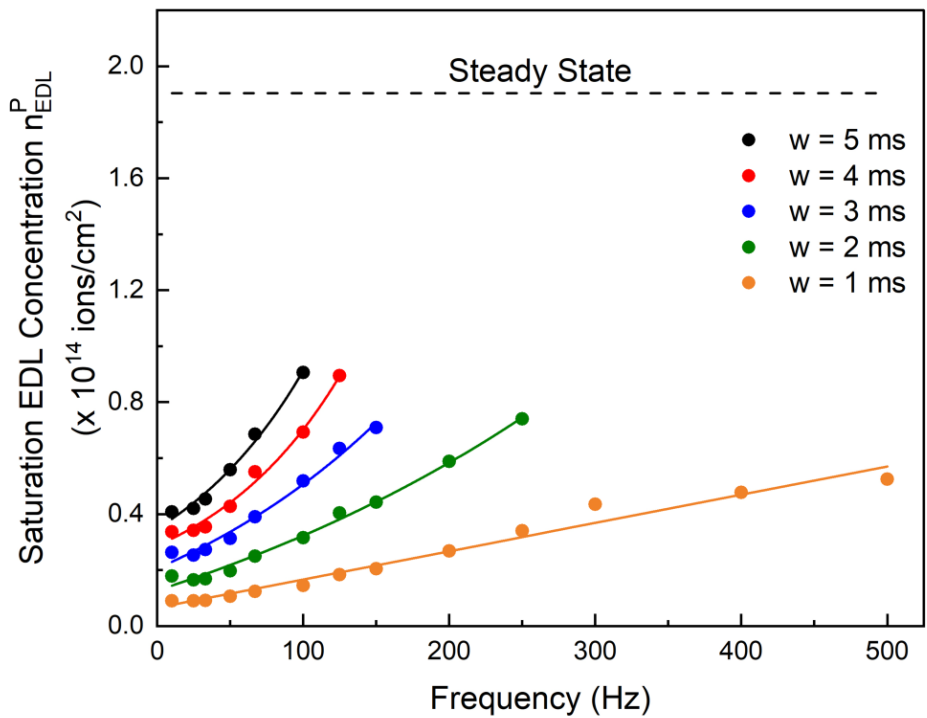

*Figure S2: Effect of frequency on saturation EDL concentration at pulse widths ranging from 1 ms to 5 ms, with steady state EDL concentration as reference.*

**S3 – LIWUS predicted results**

Using a lower starting voltage, LIWUS predicted voltages for 70 potentiation and depression pulses for the 50 nm x 50 nm capacitor geometry at a frequency of 100 Hz, as shown in Fig. S3 (a). This resulted in a $\Delta n$ of 3.06 x $10^{14}$ ions/cm$^2$ (Fig. S3 (b)), roughly 5 times larger than that produced by fixed magnitude pulses. The input voltages were only simulated for 70 pulses as the electrochemical window of 3 V was reached, but more pulses could be simulated with a lower potentiation and depression rate. Similar to the 50 nm x 50 nm geometry, LIWUS was trained with the voltage dependent steady state EDL rate data from the 1 µm x 1 µm geometry (Fig. S3 (c)) and an example linear weight update simulated in COMSOL from the LIWUS predicted voltages are shown in Fig. S3 (d).

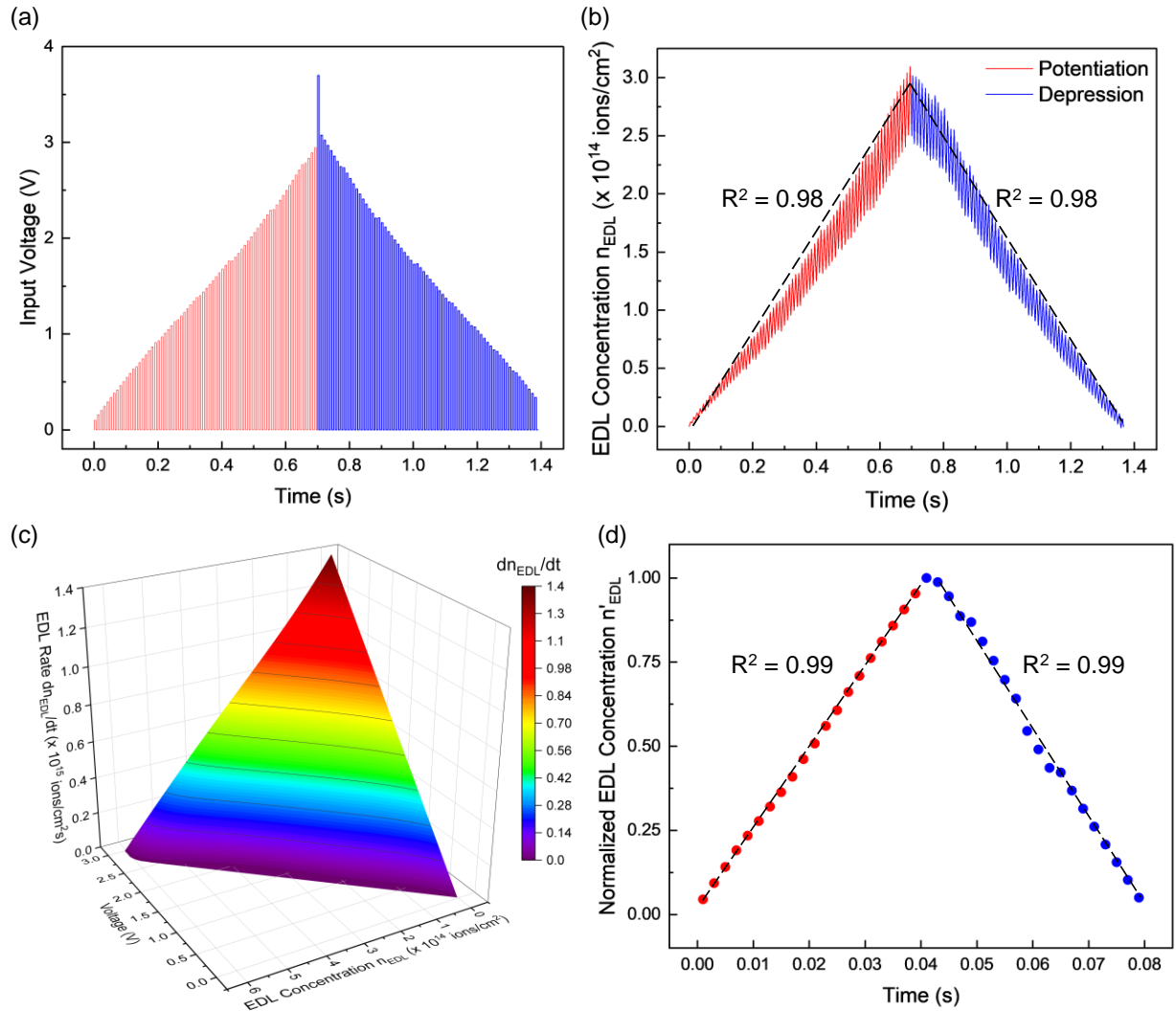

*Figure S3: (a) LIWUS predicted voltages for 70 pulses and (b) its corresponding response simulated in COMSOL with the 50 nm x 50 nm capacitor. (c) 3D plot of voltage and concentration dependent EDL rates extracted from 1 μm x 1 μm capacitor used to train LIWUS and (d) an example linear response from COMSOL predicted by LIWUS.*

**S4 – ANN training data**

The training accuracies are depicted here, as a function of the number of neurons in the hidden layer (Fig. S4 (a)) and the training epochs (Fig. S4 (b)). As training accuracies only reflect the model's performance with data that it has already been trained on, it should not be used to evaluate the true performance of the model or compare different models. However, the early stopping function used allows comparing the training efficiency of the models if not the actual accuracy values. It can be seen that the training accuracy with linear weight updates reaches 100% in 12 epochs, while the nonlinear model is still trying to improve its accuracy; this is not surprising as with supervised learning, the training accuracy reflects the model's ability to classify images it has already seen during training. Similar results can be observed as the hidden neurons are varied.

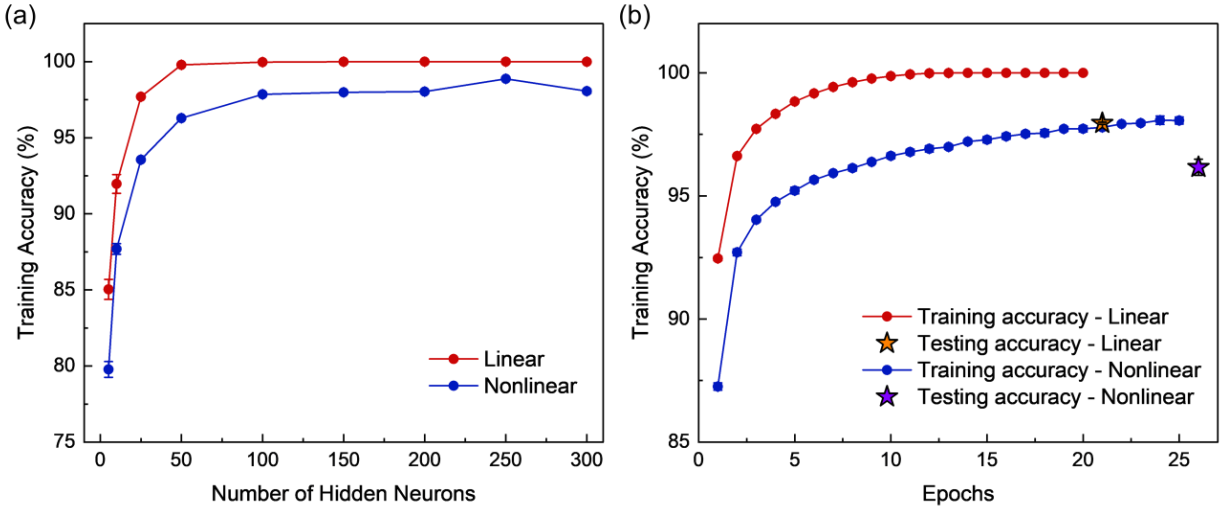

*Figure S4: Training accuracies vs. (a) hidden neurons and (b) training epochs. Testing accuracies (stars) are also shown for reference.*

**S5 – ANN image prediction accuracy**

In addition to testing the image prediction accuracy of the SNN with optimally fit models (using the early stopping function), the prediction accuracies after just one training epoch were also extracted. Of the 5000 images tested, nonlinear weight update classified 88.08% (4404 images) correctly while the linear weight update model classified 94.66% (4733 images) correctly. The labels in each box shown in Fig. S5 correspond to the number of images the SNN assigned to each of the labels. For example, with the input image as digit '2', the nonlinear weight update model classified 343 images correctly and predicted 97 images as the digit '8'. For the same digit '2', the linear weight update model classified 476 (of 500) images correctly and predicted only 7 images as the digit '8'.

# Supporting Information

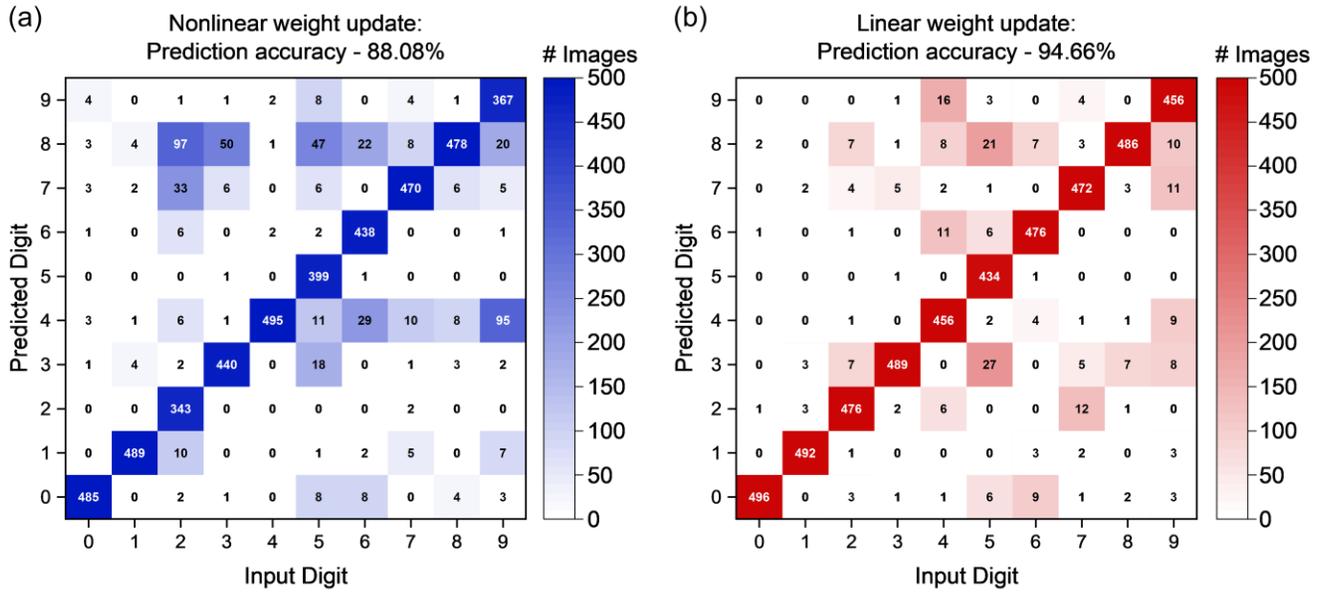

*Figure S5: Confusion matrices showing the number of digits predicted correct and wrong by the SNN against the actual input digit, after just one training epoch, for nonlinear and linear weight updates respectively.*